\documentclass{aa}  

\usepackage{graphicx}
\usepackage{txfonts}
\usepackage{lipsum}
\usepackage{subcaption}       
\usepackage{lscape}            
\usepackage{placeins}          
\usepackage{natbib}
\bibpunct{(}{)}{;}{a}{}{,}
\usepackage{hyperref}
\usepackage{threeparttablex}
\usepackage{afterpage}

\begin{document}

\title{Open cluster members in APOGEE DR17}

   \subtitle{I. Dynamics and star members}

   \author{{R. Guer\c{c}o}\inst{1,2}
        \and D. Souto\inst{3}
        \and J. G. Fern\'andez-Trincado\inst{2}
        \and S. Daflon\inst{1}
        \and K. Cunha\inst{4,1}
        \and J. V. Sales-Silva\inst{1}
        \and V. Loaiza-Tacuri\inst{1,3}
        \and V. V. Smith\inst{5}
        \and M. Ortigoza-Urdaneta\inst{6}
        \and M. P. Roriz\inst{1}
        }

   \institute{Observat\'orio Nacional / MCTI, R. Gen. Jos\'e Cristino, 77,  20921-400, Rio de Janeiro, Brazil \\
             \and Instituto de Astronom\'ia, Universidad Cat\'olica del Norte, Av. Angamos 0610, Antofagasta, Chile \\
             \email{rafael.fraga@ucn.cl, jose.fernandez@ucn.cl}
             \and Departamento de F\'isica, Universidade Federal de Sergipe, Av. Marcelo Deda Chagas, S/N Cep 49.107-230, S\~ao Crist\'ov\~ao, SE, Brazil \\
             \and Steward Observatory, University of Arizona, 933 North Cherry Avenue, Tucson, AZ 85721-0065, USA \\
             \and NSF’s NOIRLab, 950 N. Cherry Ave. Tucson, AZ 85719, USA \\
             \and Departamento de Matem\'atica, Facultad de Ingenier\'ia, Universidad de Atacama, Copiap\'o, Chile \\
             }

   \date{Received 4 April 2025 / Accepted 17 June 2025}

\abstract
{Open clusters (OCs) are groups of stars formed from the same cloud of gas and cosmic dust. They play an important role in studies of star formation and evolution and our understanding of galaxy structure and dynamics.}
{The main objective of this work is to identify stars that belong to OCs using astrometric data from Gaia EDR3 and spectroscopic data from APOGEE DR17. Furthermore, we investigate the metallicity gradients and orbital properties of the OCs in our sample.}
{By applying the HDBSCAN clustering algorithm to these data, we identified observed stars in our galaxy with similar dynamics, chemical compositions, and ages.
The orbits of the OCs were also calculated using the GravPot16 code.}
{We find 1,987 stars that tentatively belong to 49 OCs; 941 of these stars have probabilities above 80 \% of belonging to OCs. Our metallicity gradient presents a two-slope shape for two measures of different Galactic center distances -- the projected Galactocentric distance and the guiding center radius to the Galactic center -- as already reported in previous work. However, when we separate the OCs by age, we observe no significant difference in the metallicity gradient slope beyond a certain distance from the Galactic center. Our results show a shallower gradient for clusters younger than 2 Gyr than those older than 2 Gyr. All our OCs dynamically assemble the disk-like population very well, and they are in prograde orbits, which is typical for disk-like populations. Some OCs resonate with the Galactic bar at the Lagrange points L4 and L5.}
{}


\maketitle

\section{Introduction}

Open clusters (OCs) are composed of groups of stars that are believed to have formed from the same molecular cloud in a relatively short time. Their stellar members thus share a nearly common age and chemical composition, making them excellent laboratories for studying stellar evolution.  Milky Way OCs are predominantly located along the Galactic plane (GP), with the youngest examples typically found within 100 pc of the GP. In contrast, older clusters can be located at Galactic altitudes greater than 1 kpc from the plane \citep{Cantat2024}.  Due to their coeval nature and nearly identical chemistry, cluster members can be fit by stellar isochrones, resulting in distances that are relatively well constrained, and OC chemical abundances have played a pivotal role in probing chemical trends, such as abundance gradients, across the Galactic disk \citep{Casamiquela2021}.  In addition to
their relatively well-defined distances, the ages of OCs are also known and have been used to study chemical evolution and abundance gradients as functions of time. 
Recent advances in observational technology, including larger telescopes, multi-fiber spectrographs, and large high-resolution spectroscopic surveys, such as Gaia-ESO Public Spectroscopic Survey \citep[Gaia-ESO;][]{Gilmore2012}, Galactic Archaeology with HERMES\footnote{HERMES is a multi-fibre spectrograph.} \citep[GALAH;][]{Martell2017}, and Apache Point Observatory Galactic Evolution Experiment \citep[APOGEE;][]{Majewski2017}, have significantly broadened our understanding of OCs, while the Gaia mission \citep{Gaia2016,Gaia2018,Gaia2021} has been fundamental in expanding the known census of OCs to nearly 7,000 \citep[][]{Hunt2023}.  Maximizing the scientific return offered by OC stars requires, as a first step, identifying bona fide cluster members with minimal contamination from interloping field-star nonmembers. Techniques for determining stellar membership in clusters have advanced in recent years through the use of unsupervised machine learning methods, particularly density-based hierarchical algorithms, as illustrated by \cite{Castro2018}, \cite{Cantat2020}, and \cite{Castro2022}.

As reported by \cite{Randich2022}, the Gaia-ESO Survey (GES), carried out over a decade, has delivered a final spectroscopic catalog comprising more than 100,000 stars and includes advanced stellar parameters, chemical abundances, and cluster membership information. The GES is the only extensive spectroscopic survey conducted with an 8-meter-class telescope to date, and it was carried out using a multi-pipeline analysis strategy and a refined homogenization procedure, establishing methodological standards that will be adopted by upcoming surveys such as WHT\footnote{William Herschel Telescope} Enhanced Area Velocity Explorer \citep[WEAVE;][]{Dalton2016} and 4-metre Multi-Object Spectroscopic Telescope \citep[4MOST;][]{deJong2019}. The final data products, especially when combined with Gaia (Early) Data Release 3 ((E)DR3) and asteroseismic data, are expected to provide a long-term legacy for Galactic structure and evolution studies.

In the era of big data, machine learning has become crucial for obtaining valuable insights and making data-driven decisions. Machine learning refers to the ability of software to learn from data and improve autonomously from statistical algorithms without the need for explicit programming for each specific task. This field encompasses various techniques and algorithms designed to recognize patterns, make predictions, and efficiently classify data \citep{Hopfield1986}. 
Within the broad scope of machine learning, clustering techniques play a key role in organizing datasets into homogeneous groups. Among the most advanced and flexible techniques in this domain is the Hierarchical Density-Based Spatial Clustering of Applications with Noise \citep[HDBSCAN;][]{Campello2013,Campelo2015}. HDBSCAN extends the widely used Density-Based Spatial Clustering of Applications with Noise (DBSCAN) algorithm \citep{Ester1996} with a greater capacity to identify structures in high-density data and effectively manage noise. This enables the identification of data clusters of varying sizes and shapes without needing a prior specification of the number of clusters. In the works of \cite{Hunt2021,Hunt2023}, \cite{Chi2023}, \cite{Locaiza2023}, and \cite{Grilo2024}, HDBSCAN has shown its effectiveness in isolating high-probability stellar members of OCs. 

The dynamical evolution of OCs is shaped by tidal forces from the Galactic disk and molecular clouds \citep{Friel1995}. Theoretical models predict that OCs generally dissolve within 10$^8$ to 10$^9$ years, with disruption timescales depending on their mass and core radius \citep{Spitzer1958,Spitzer1958a}. Only the most massive and centrally concentrated clusters, or those in less disruptive orbits, tend to survive \citep{Friel1995}. The motion and spatial distribution of Galactic OCs provide valuable insight into the gravitational potential and perturbations that influence the structure and dynamics of the Milky Way \citep{Soubiran2018}.
Computational tools have been developed to model the Galactic gravitational potential and therefore study its dynamics: NEMO \citep{Teuben1995}, GalPot \citep{Dehnen&Binney1998,McMillan2017a,McMillan2017b}, galpy \citep{Bovy2015}, GravPot16 \citep{Fernandez-Trincado2017} and AGAMA \citep{Vasiliev2019}. This work uses the GravPot16 code, a tool designed for Galactic dynamics and orbit integration, whose steady-state gravitational potential has been inferred from the superposition of the multiple galactic components of the popular Besan\c{c}on galaxy model \citep{Robin2003,Robin2012,Robin2014}. It enables detailed simulations of the OC dynamics, providing insights into their past trajectories and future evolution.
By integrating observational data with sophisticated models such as GravPot16, researchers can reconstruct the orbits of OCs \citep{Schiappacasse-Ulloa2018}, assess their interactions with Galactic structures, and assess their susceptibility to disruption.

This paper is organized as follows: The methodology for obtaining OC stellar members that were observed spectroscopically by the APOGEE survey is presented in Section \ref{sec:metodology}.  The OC members identified here and their cluster membership probabilities are compared with those from other studies in Section \ref{sec:membership}.  Section \ref{sec:discussion} provides a discussion of the dynamics of the APOGEE OC sample, along with the metallicity (as [Fe/H]) gradients derived from these clusters.  Conclusions are then presented in Section \ref{sec:conclusions}.

\section{Methodology and sample} \label{sec:metodology}

\subsection{HDBSCAN} \label{sec:hdbscan}

We applied the HDBSCAN machine learning algorithm to a sample of stars observed by Gaia and the APOGEE survey to identify stars with similar dynamics, chemical compositions, and ages in our Galaxy, specifically those belonging to OCs. 
HDBSCAN builds on DBSCAN by using the minimum number of points (m$_{Pts}$) parameter to define core and border points and introduces the minimum cluster size (m$_{clSize}$) parameter, which sets the minimum number of data points needed to form a cluster. Fine-tuning the m$_{clSize}$ parameter is crucial in optimizing HDBSCAN's performance, as it depends on the dataset's characteristics. 
A larger m$_{clSize}$ leads to fewer and larger clusters, potentially merging smaller, denser clusters into larger ones. In contrast, a smaller m$_{clSize}$ identifies smaller, denser clusters but may also result in more noise being classified as small clusters. The HDBSCAN hierarchy reflects clusters that appear and merge at multiple density levels. The stability of clusters is the key metric for the selection of clusters, with broad bases and sharp density peaks considered the most meaningful. This hierarchical approach makes HDBSCAN ideal for complex datasets, such as stellar populations, where structures at different density levels can exist. 

\begin{figure}
\centering
\includegraphics[width=\columnwidth]{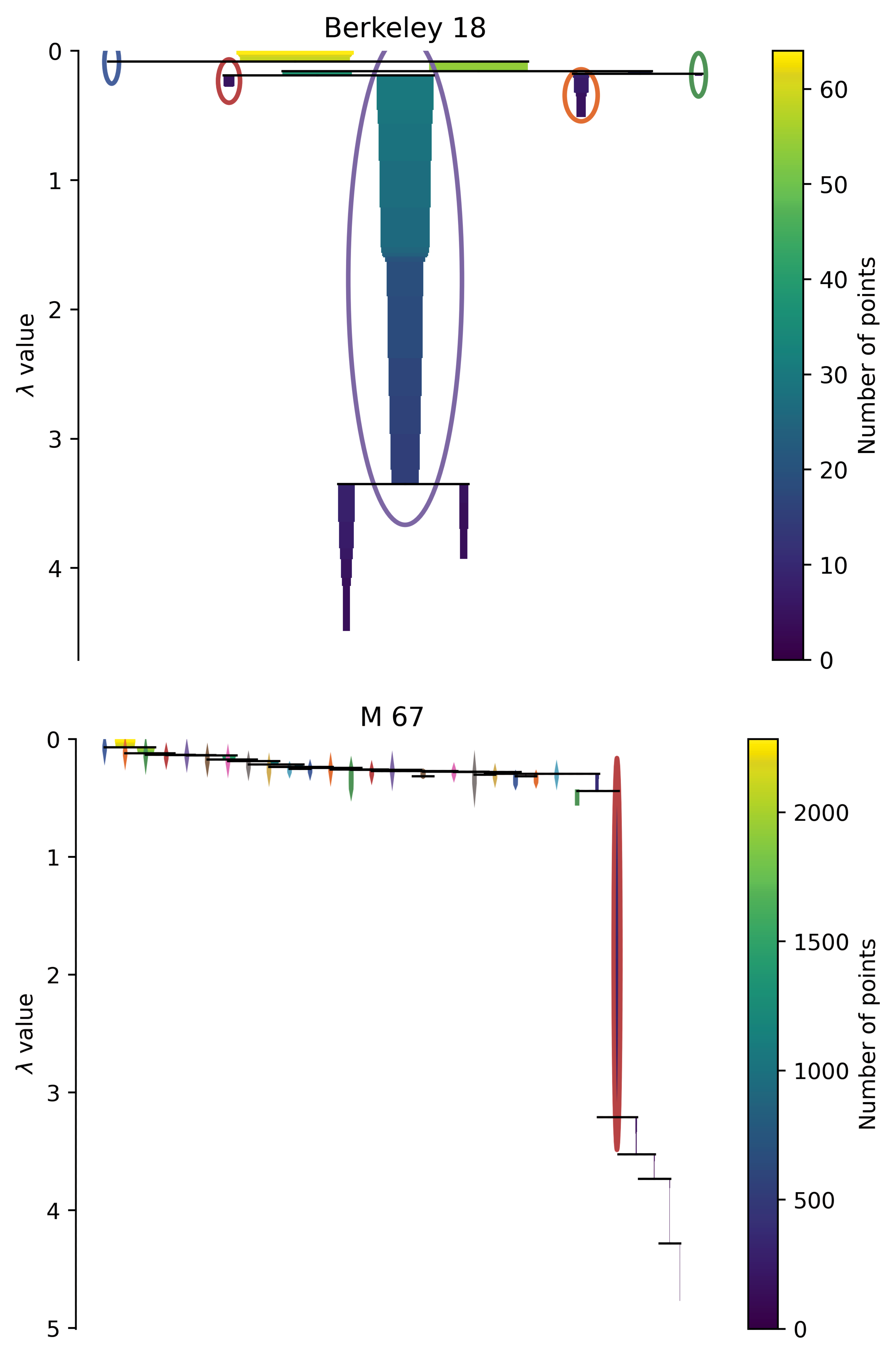}
\caption{Cluster tree, or dendrogram, produced by HDBSCAN code. The $\lambda$ value, being a measure related to the density-based nature of the HDBSCAN, is equal to the inverse of the core distance and helps in understanding the hierarchy and persistence of clusters.}
\label{fig:hdbscan}
\end{figure}

We adopted the same values for m$_{Pts}$ and m$_{clSize}$ as recommended by \cite{Campello2013}. Using equal values for m$_{Pts}$ and m$_{clSize}$ ensures consistency between the density definition and cluster size threshold in HDBSCAN. This approach avoids introducing biases by keeping the clustering process uniform and interpretable since the minimum cluster size directly aligns with the density required to form a cluster. It also simplifies parameter tuning while maintaining robust results, especially when working with datasets where clusters of varying densities coexist, fostering a more reliable identification of meaningful structures in the data. In Figure \ref{fig:hdbscan}, we present the HDBSCAN cluster tree for two OCs confirmed in this work, Berkeley 18 and M67. The cluster tree visually represents the hierarchical structure of the sample, illustrating how the clusters are nested within each other and how they merge as the distance threshold increases. The $\lambda$ value in Figure \ref{fig:hdbscan} is the inverse of the core distance, where the core distance is defined as the distance to the k$^{th}$ nearest neighbor, with k being a user-specified parameter. Higher values of $\lambda$ correspond to higher density levels (more densely packed points). The height of a branch in the cluster tree (measured by $\lambda$) reflects the density level at which clusters merge or split. A higher branch height indicates that the cluster remains stable at higher densities, and clusters that persist over a wide range of $\lambda$ values (have long branches) are considered more stable and robust. When moving up the tree (increasing $\lambda$), we see how smaller clusters merge into larger ones, indicating a hierarchical structure. The value of $\lambda$ helps identify the most stable clusters; by examining the range of $\lambda$ over which a cluster exists, one can determine its persistence and robustness. Clusters with high $\lambda$ stability scores are considered meaningful. Points or clusters with shallow $\lambda$ values are often considered noise or outliers since they only exist at low-density levels. We can then assume that the branches with greater values of $\lambda$ in Figure \ref{fig:hdbscan} represent the members of the OCs Berkeley 18 (top panel) and M 67 (bottom panel).

\subsection{Sample}

Our analysis is based on the final results of Sloan Digital Sky Survey IV (SDSS-IV) APOGEE Data Release 17 (DR17), as reported by \cite{Abdurro2022}. APOGEE spectra are obtained using two cryogenic, multi-fiber spectrographs (with 300 fibers) that observe in the near-infrared H-band, covering wavelengths from 1.51 to 1.69 $\mu$m  \citep{Gunn2006,Wilson2010,Wilson2019}. These spectrographs are located in the northern and southern hemispheres: one at Apache Point Observatory (APO) in New Mexico, USA, and the other at Las Campanas Observatory (LCO) near La Serena, Chile.
The main objective of the APOGEE survey was to determine the chemical composition and kinematics of red giant stars to study the chemical evolution and dynamics of stellar populations in the Milky Way.  A large number of OC stars were observed and analyzed as part of this effort.

The APOGEE DR17 dataset contains more than 657,000 stars, with stellar parameters and chemical compositions derived from high-resolution spectra ($R \approx$ 22,500). Radial velocities (RVs) and metallicities for all stars were automatically determined by the APOGEE pipeline \citep{Nidever2015,GarciaPerez2016}, with typical uncertainties of 1.0 km$\cdot$s$^{-1}$ for RV and approximately 0.02 dex for metallicity. 

\section{Membership analysis from HDBSCAN} \label{sec:membership}

As a first step, we selected OCs from DR17 based on the right ascension (RA) and declination (Dec.) of their centers as given in previous studies (\citealt{Donor2020}, OCCAM; \citealt{Cantat2020, Hunt2023}). 
We applied filters to these data to ensure quality, including removing possible binaries from the sample. The APOGEE VSCATTER parameter, which measures the variation in RV in multiple observations, was used as an indicator of binarity, with a cutoff threshold of 1 km$\cdot$s$^{-1}$, a typical value adopted by other studies \citep{Bailer-Jones2021,Quispe2022}, and validated, for example, by \cite{Jonsson2020}. 

The following parameters were used as input for HDBSCAN in our analysis: RA, Dec., parallax ($\pi$), proper motions ($\mu_{RA}$ and $\mu_{Dec}$), RV, and metallicity ([Fe/H]). The parameters RA, Dec., $\pi$, $\mu_{RA}$, and $\mu_{Dec}$ are sourced from the Gaia (E)DR3 data \citep[][]{Gaia2021}, while RV and [Fe/H] are taken from the APOGEE DR17 dataset \citep[][]{Abdurro2022}. 
The region selected to be considered in the HDBSCAN analysis for each OC was determined using the cluster center. 
From the center of each OC, we defined the ratio between the window size in RA or Dec. ($\delta$(RA) and $\delta$(Dec)) and the GAIA DR3 parallax of each OC as 6. This ratio was empirically chosen and has been well suited to improve our membership analysis.
Nonetheless, in a few cases, we considered this ratio with values different from 6 -- Berkeley 98 (ratio 8) and Berkeley 19, Berkeley 33, and King 8 (ratio 10) -- for better operation of the HDBSCAN code. 

\subsection{Cluster membership}

We identified 1,987 stars across 49 OCs, with 1,456 stars having a membership probability above 50\% and 941 stars with a probability above 80\%. These results are presented in Tables \ref{tab:stars} and \ref{tab:parameters}. HDBSCAN assigns a membership probability to each star by evaluating its proximity to the densest regions of the OC across the combined dimensions. Stars that exhibit consistent spatial location, velocity, and chemical composition with the core members receive higher probabilities (close to 1). Conversely, stars that deviate in one or more parameters receive lower probabilities, reflecting their possible outlier or transition status. This approach ensures a robust classification of the OC membership by accounting for kinematic and metallicity coherence among stars. 

We ran HDBSCAN iteratively, varying m$_{clSize}$ from 2 to 10, and computed the mean and standard deviation of the input parameters for each run. The optimal m$_{clSize}$ value for each OC was determined by selecting the value that led to the smallest standard deviation of the 
mean metallicity of stars with membership probabilities greater than 80\%. 
This assumes, of course, that OCs have homogeneous chemical abundances.
We note that in cases where the standard deviation was approximately equal for multiple m$_{clSize}$ values, we used the smallest standard deviation from the metallicities including all stars from the OC group selected by HDBSCAN.

\begin{figure}
\centering
\includegraphics[width=\columnwidth]{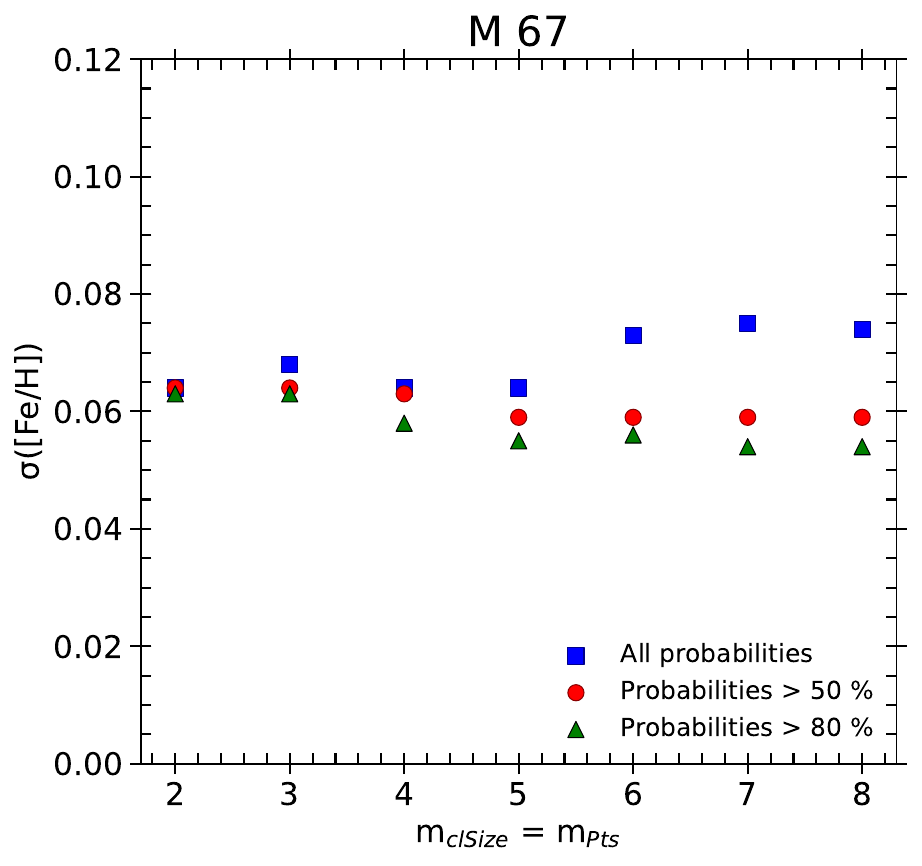}
\caption{m$_{clSize}$ vs. $\sigma$([Fe/H]) of M 67. The blue squares represent all stars, the red circles stars with probabilities greater than 50 \%, and the green triangles stars with probabilities greater than 80 \%.}
\label{fig:min_met}
\end{figure}

Figure \ref{fig:min_met} illustrates how m${clSize}$ was selected for the OC M 67. For each m${clSize}$, the standard deviation of the mean 
metallicity ($\sigma$([Fe/H])) was calculated for all stars (blue squares), stars with probabilities greater than 50\% (red circles), and stars with probabilities greater than 80\% (green triangles). In the case of M 67, we selected m${clSize}$ = 5, as it resulted in one of the lowest $\sigma$([Fe/H]) for stars with membership probabilities greater than 80\%. In this case, we do not consider m${clSize}$ = 7, with $\sigma$([Fe/H]) approximately equal to m${clSize}$ = 5 for stars with membership probabilities greater than 80\%, because the $\sigma$([Fe/H]) for all stars (blue squares) is much greater than $\sigma$([Fe/H]) obtained with m${clSize}$ = 5. The m${clSize}$ values for all 49 OCs analyzed in this study are provided in Table \ref{tab:parameters}. 

\begin{figure}
\centering
\includegraphics[width=\columnwidth]{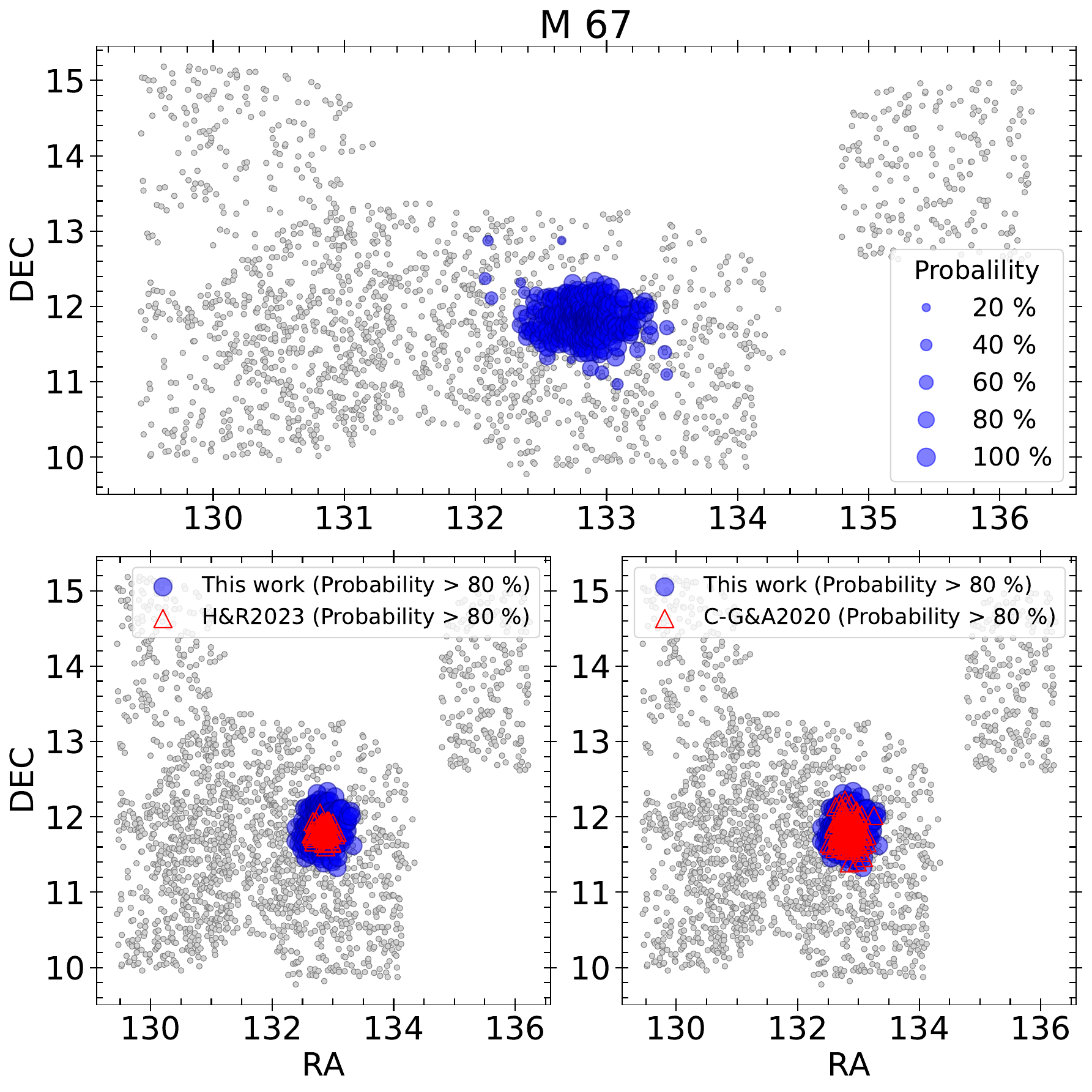}
\caption{Upper panel: Window RA vs. Dec. centered on the position of M 67 by \cite{Donor2020}, \cite{Cantat2020} and \cite{Hunt2023}. The gray circles are the stars used in HDBSCAN to search for members of M 67. The blue circles are the members obtained with the HDBSCAN code. The caption indicates the probability of each star belonging to the OC according to the size of the blue circle. Lower-left panel: Stars with a probability of belonging to the OC greater than 80 \% according to our results (blue circles) and the results of \citet[][red triangles]{Hunt2023}. Lower-right panel: Stars with a probability of belonging to the OC greater than 80 \% according to our results (blue circles) and the results of \citet[][red triangles]{Cantat2020}.}
\label{fig:ra_dec}
\end{figure}

The HDBSCAN clustering result for M 67 is shown in the top panel of Figure \ref{fig:ra_dec}. We chose a search window for M67 members with 2,283 stars (gray circles) centered on the central cluster position obtained in \cite{Donor2020}, \cite{Cantat2020}, and \cite{Hunt2023}. The stars selected as members are represented with blue circles, whose sizes 
represent the probability of belonging to the OC (378 stars for M 67). The number of stars (N$_{all}$) obtained for each cluster is shown in Table \ref{tab:parameters}.

\subsection{Comparison with previous studies}

The lower panels of Figure \ref{fig:ra_dec} show the cross-match of our results with \citet[][red triangles in the lower-left panel]{Hunt2023} and \citet[][red triangles in the lower-right panel]{Cantat2020} for M67 stars. In this figure, we show only stars with probabilities above 80 \% from \cite{Hunt2023} and \cite{Cantat2020}. In this example, we identified 257 stars with probabilities greater than 80 \% of belonging to M 67. Compared with the same probability range (P$>$80\%), we have 84 stars in common with \cite{Hunt2023} and 146 stars in common with \cite{Cantat2020}. 
For NGC 2158, for example, we have 27 stars in common with \cite{Hunt2023} and 54 stars in common with \cite{Cantat2020}, for stars with probabilities greater than 80 \% of being members of the cluster.

\begin{figure}
\centering
\includegraphics[width=\columnwidth]{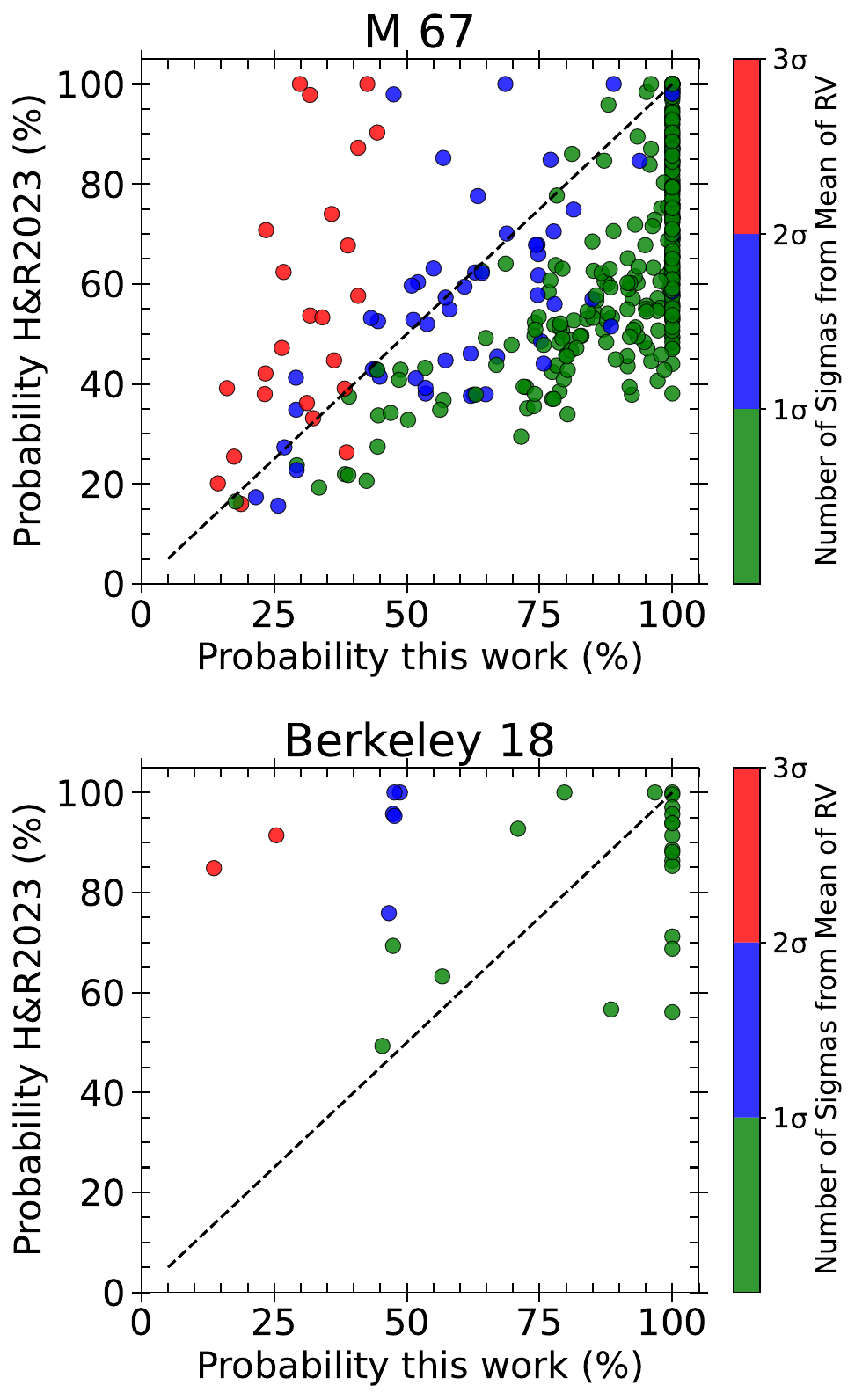}
\caption{Probabilities from this work vs. probabilities from \cite{Hunt2023} (H\&R2023) for M 67 (top panel) and Berkeley 18 (lower panel) stars. The color bar shows the number of standard deviations ($\sigma$) from the cluster RV mean: $\leq$ 1$\sigma$ (green), between 1$\sigma$ and 2$\sigma$ (blue), and between 2$\sigma$ and 3$\sigma$ (red).}
\label{fig:probability}
\end{figure}

A comparison of our membership results with \cite{Hunt2023}, which employed the HDBSCAN code using the astrometric parameters from GAIA DR3 of RA, Dec., $\mu_{RA}$, $\mu_{Dec}$ and $\pi$ as input parameters, finds that we have more stars classified by HDBSCAN as members with probabilities above 80 \% when compared with \cite{Hunt2023}. 
The probable explanation for this is illustrated in Figure \ref{fig:probability}, which compares our membership probabilities with those from \cite{Hunt2023} for M 67 (top panel) and Berkeley 18 (lower panel) stars. Our membership probability is thus weighted by RV, as we have adopted this as an input parameter, unlike the analysis of \cite{Hunt2023}.

Regarding metallicities, we find that OCs exhibit [Fe/H] values with low $\sigma$ for their stars that have probabilities greater than 80\%.  For the 49 OCs included here, the typical value of $\sigma$([Fe/H])$\sim$0.04 dex, using the [Fe/H] values from DR17.  These standard deviations were used as a criterion in HDBSCAN to determine the value of m$_{clSize}$ for each OC (see Table \ref{tab:parameters}). 
Furthermore, Figure \ref{fig:probability} highlights the importance of including RV as a criterion to enhance confidence in stars classified with probabilities of being OC members in \cite{Hunt2023}. Stars with RV values closer to the cluster average tend to show higher membership probabilities, with many achieving 100\%. This pattern is also consistently observed in other OCs. 
For example, stars with RV standard deviations between $2\sigma$ and $3\sigma$ have probabilities below 50\% of being M 67 members (top panel of Figure \ref{fig:probability}) or Berkeley 18 members (bottom panel of Figure \ref{fig:probability}).
These low $\sigma$ values obtained align with recent works on OC homogeneity using the APOGEE spectra, such as \citet{Bovy2016} and \citet{Sinha2024}. The authors find limits on OC homogeneity within 0.02 dex or less for most elements.

\begin{figure*}
\centering
\includegraphics[width=0.95\textwidth]{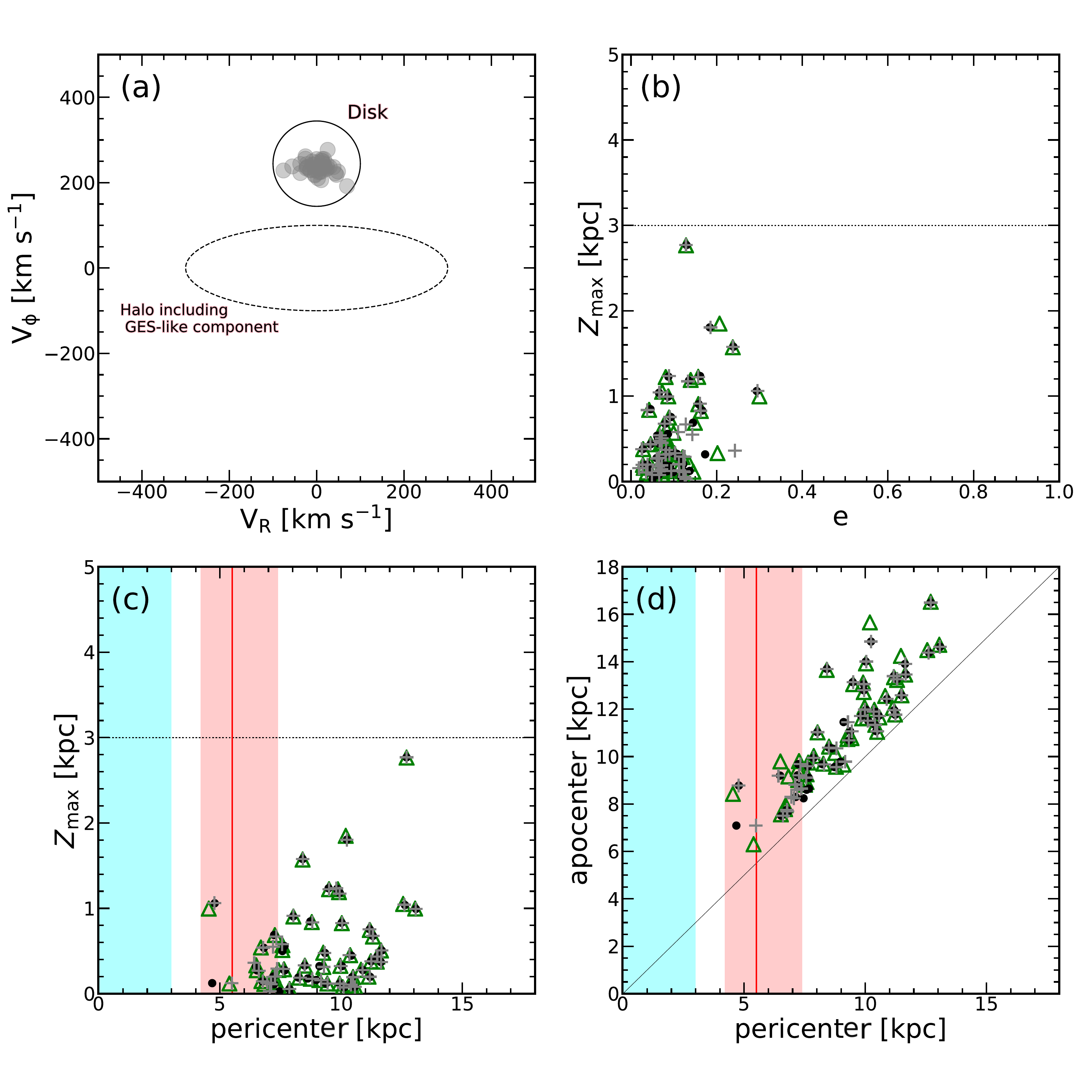}
\caption{Kinematics and orbital elements of selected OCs. Panel (a): Galactocentric RV (V$_{\rm R}$) vs. azimuthal velocity (V$_{\rm \phi}$) for 49 selected OCs (gray symbols) that fall within the area occupied by the disk population (black circle). For guidance, a dashed black ellipsoid also highlights the halo Gaia-Enceladus-Sausage-like \citep{Belokurov2018} area. Remaining panels: Orbital elements assuming $\Omega_{\rm bar}=$ 31 (green empty symbols), 41 (black symbols), and 51 (gray crosses) km s$^{-1}$ kpc$^{-1}$. The horizontal dotted black lines indicate a distance of $\sim 3$ kpc \citep[i.e., the edge Z$_{\rm max}$ of the thick disk;][]{Carollo2010}, while the cyan and red shaded regions indicate the radius \citep[i.e., 3 kpcs;][]{Barbuy2018} of the Milky Way bulge and location of the bar's CR between 4.2 kpcs ($\Omega_{\rm bar} = $ 51 km s$^{-1}$ kpc$^{-1}$)  and 7.4 kpcs ($\Omega_{\rm bar} = $ 31 km s$^{-1}$ kpc$^{-1}$), with the vertical thick red line indicating the position of 5.5 kpc for $\Omega_{\rm bar} = $ 41 km s$^{-1}$ kpc$^{-1}$. The straight line in panel (d) shows the one-to-one line, which indicates that a cluster on this line would have a circular orbit.}
\label{fig:dynamical}
\end{figure*}

\section{Discussion} \label{sec:discussion}

\subsection{Orbital elements of selected OCs}

We used the state-of-the-art Milky Way model  GravPot16\footnote{\url{https://gravpot.utinam.cnrs.fr}} to predict the orbital path of OCs from our sample 
in a steady-state gravitational Galactic model that includes a ``boxy/peanut" bar structure \citep{Fernandez-Trincado2019}.  
This Galactic model is fundamental for predicting the structural and dynamic parameters of the Galaxy to the best of our recent knowledge of the Milky Way. 
We adopted the same model configuration, solar position, and velocity vector as described in \citet{Fernandez-Trincado2019} for the orbit computations, except for the angular velocity of the bar ($\Omega_{\rm bar}$), for which we used the recommended value of 41 km s$^{-1}$ kpc$^{-1}$ \citep[][]{Sanders2019}, and assuming variations of $\pm$10 km s$^{-1}$ kpc$^{-1}$. 
These structural parameters for our bar model (e.g., mass and orientation) are within observational estimations that lie in the range of 1.1$\times$10$^{10}$ M$_{\odot}$ and present-day orientation of 20$^{\circ}$ \citep[value adopted from dynamical constraints, as highlighted in Fig. 12 of][]{Tang2018} in the non-inertial frame (where the bar is at rest). The lengths of the bar scale are $x_0=$ 1.46 kpc, $y_{0}=$ 0.49 kpc, and $z_0=$0.39 kpc, and the middle region ends on the effective semimajor axis of the bar at $Rc = 3.28$ kpc \citep{Robin2012}. 

For guidance, the Galactic convention adopted in this work is: $X-$axis is oriented toward $l=$ 0$^{\circ}$ and $b=$ 0$^{\circ}$, $Y-$axis is oriented toward $l$ = 90$^{\circ}$ and $b=$0$^{\circ}$, and the disk rotates toward $l=$ 90$^{\circ}$; the velocity is also oriented in these directions. Following this convention, the Sun's orbital velocity vectors are [U$_{\odot}$, V$_{\odot}$, W$_{\odot}$] = [$11.1$, $12.24$, 7.25] km s$^{-1}$, respectively \citep{Ralph2010}. The model has been rescaled to the Sun's Galactocentric distance, 8.27 kpc \citep{GRAVITY2021}, and a circular velocity at the solar position of $\sim$ $229$ km s$^{-1}$ \citep{Eilers2019}.

The most likely orbital parameters and their uncertainties are estimated using a simple Monte Carlo scheme. An ensemble of a half million orbits was computed backward in time for 2 Gyr, under variations of the observational parameters assuming a normal distribution for the uncertainties of the input parameters (e.g., positions, heliocentric distances, RVs, and proper motions), which were propagated as 1$\sigma$ variations in a Gaussian Monte Carlo resampling. To compute the orbits, we adopted a mean RV (Table \ref{tab:parameters}) of the member stars computed from the APOGEE DR17 data \citep[][]{Abdurro2022}. The nominal proper motions ($<\mu_{RA}>$ and $<\mu_{Dec}>$ from Table \ref{tab:parameters}) for each cluster were taken from Gaia EDR3 \citep[][]{Gaia2021}, with an assumed uncertainty of 0.5 mas yr$^{-1}$ for the orbit computations. Heliocentric distances ($d_{\odot}$) were adopted from \cite{Bailer-Jones2021}. The results for the main orbital elements are listed in Table \ref{tab:dynamics}. 

Figure \ref{fig:dynamical} shows a scatter plot of all possible combinations among the orbital elements in the non-axisymmetric model. In the same figure, we use three different angular velocities for the Galactic bar. All selected clusters share the same kinematical behavior as the disk population, as illustrated in Fig. \ref{fig:dynamical}(a). Orbital elements reveal that most clusters have low vertical excursions ($Z_{\rm max} <$ 2 kpc) from the GP, except for Berkeley~20 (among the old clusters in our sample) that reach distances ($\sim$ 2.76$\pm$0.84 kpc) closest to the edge Z$_{\rm max} \sim 3$ kpc of the thick disk \citep[e.g.,][]{Carollo2010}. It is not surprising that Berkeley~20 exhibits such high excursions in the GP, as we know that this cluster belongs to the old OCs in this sample, i.e., most of the old (AGE $>$ log$_{10}$(8.6) Gyr) OCs in our sample exhibit a large scatter ($\sim$ 2 kpc) in $Z_{\rm max}$ with vertical excursions from the GP $Z_{\rm max}> 0.2$ kpc. In comparison, young (AGE $<$ log$_{10}$(8.6) Gyr) OCs are confined to less than $\sim$0.2 kpc within the GP. 
For eccentricity, all clusters show low eccentricities, $e<0.3$, and are characterized by prograde orbits relative to the direction of Galactic rotation. 

\begin{figure*}
\centering
\includegraphics[width=1.0\textwidth]{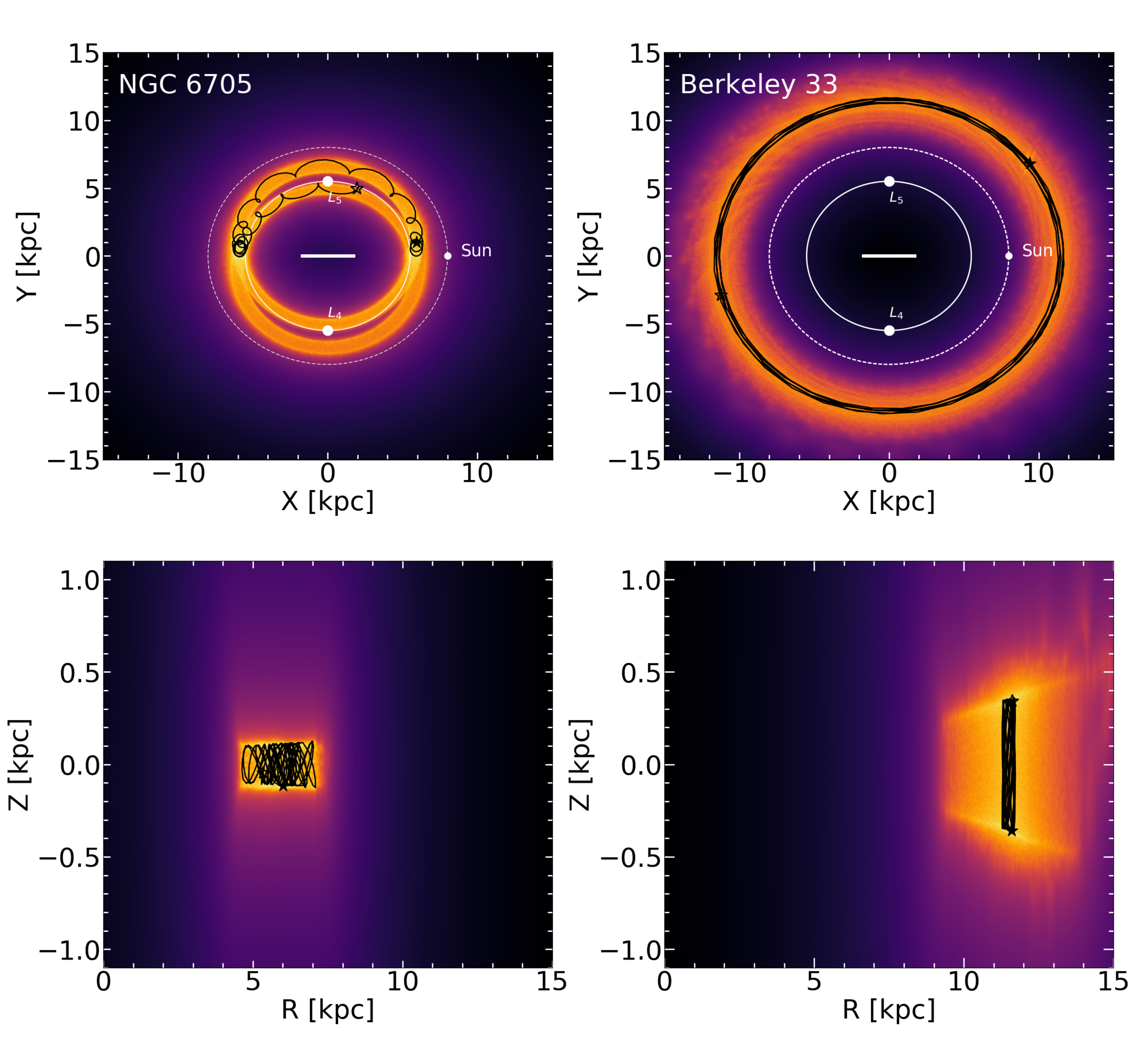}
\caption{Ensemble of half a million orbits in the frame corotating with the bar for two selected OCs, projected on the equatorial (top) and meridional (bottom) GPs in the non-inertial reference frame with a bar pattern speed of 41 kms$^{-1}$ kpc$^{-1}$, and time-integrated backward over 2 Gyr. Yellow and orange correspond to more probable regions of the space, which are most frequently crossed by the simulated orbits. The inner solid and outer dashed white circles in the top panels show the locations of the CR and solar orbit, respectively, while the white dots mark the positions of the Lagrange points of the Galactic bar, $L_4$ and $L_5$, and the current Sun's position. The horizontal solid white line shows the extension of the bar \citep[$R_c \sim3.4$ kpc;][]{Robin2012} in our model. The solid black line shows the orbits of the selected OCs from the observables without error bars. The black-filled and unfilled star symbols indicate the initial and final positions of the OCs in our simulations, respectively.}
\label{MonteCarlo}
\end{figure*}

Figure \ref{MonteCarlo} shows two selected examples of the simulated orbits by adopting a simple Monte Carlo approach. The probability densities of the resulting orbits projected on the equatorial and meridional GPs in the non-inertial reference frame, where the bar is at rest, are highlighted in the same figure. It is interesting to note that some particular cases shown in Figure \ref{MonteCarlo}, such as NGC~6705, exhibit a chaotic orbital configuration that has a banana-shaped shape parallel to the bar, which circulates the Lagrange points $L_4$ and $L_5$, with orbits that move around the corotation radius (CR) but eventually become trapped by Lagrange point $L_4$ or $L_5$. This is unusual dynamical behavior for bar-induced OCs and is likely related to some moving groups, such as the so-called Hercules stream \citep{Perez-Villegas2017}. Our results could provide insight into the formation processes and origin of the moving groups in the solar neighborhood.  

Additionally, a compilation of simulated orbital projections on the equatorial GPs in the non-inertial reference frame with a bar pattern speed of 31 km s$^{-1}$ kpc$^{-1}$, 41 km s$^{-1}$ kpc$^{-1}$ and 51 km s$^{-1}$ kpc$^{-1}$ are shown in the Appendix in Figures \ref{equatorial1}, \ref{equatorial2}, and \ref{equatorial3}, respectively. 
The orbits shown in these figures have been simulated by adopting the observable RV, heliocentric distance, and proper motions without error bars. A few OCs in disk-like orbits show oscillations in and out of the CR, meaning these OCs are trapped in resonance with the bar and moving around some of the Lagrange points. In other words, these few OCs are likely describing chaotic orbits. However, there is a strong dependence on the bar angular velocity, $\Omega_{bar}$, demonstrating that including a Galactic bar is important to describe the dynamical history of the OCs in the Galaxy.    

\subsection{Metallicity gradients}

\begin{figure}
\centering
\includegraphics[width=\columnwidth]{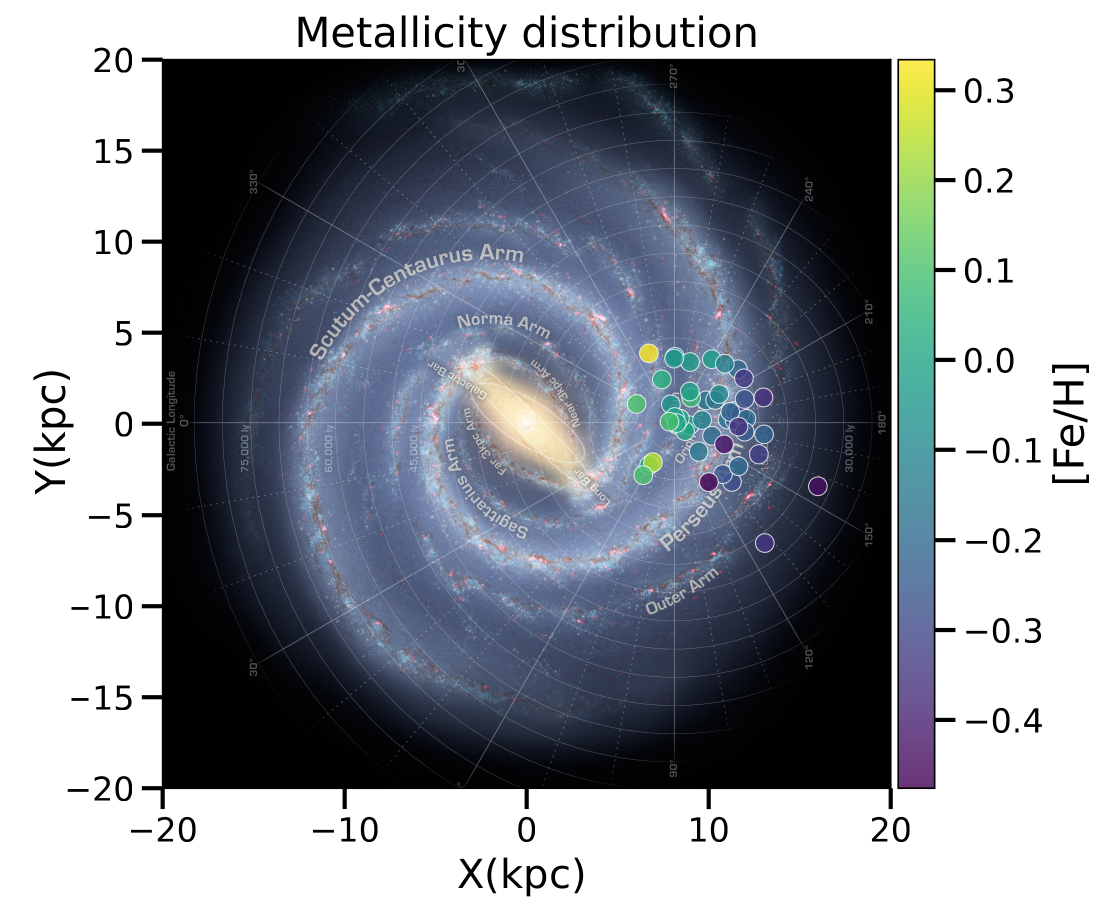}
\caption{Distributions in Galactocentric coordinations X and Y of the 49 OCs. The color bar represents the metallicity between -0.48 and 0.33 dex. We used \texttt{mw-plot} \url{https://milkyway-plot.readthedocs.io}}
\label{fig:met_xy}
\end{figure}

Figure \ref{fig:met_xy} presents the distribution of the 49 OCs studied in Galactocentric coordinates X and Y (Table \ref{tab:dynamics}). Values of [Fe/H] (Table \ref{tab:parameters}) span the interval between -0.48 and +0.33 dex. We used the mw-plot\footnote{\url{https://milkyway-plot.readthedocs.io}} code to obtain the Milk Way map that is modified from an image by NASA/JPL-Caltech/R. Hurt (SSC/Caltech). The clusters occupy distances from the Galactic center between 6.14 and 16.36 kpc, displaying the well-known anticorrelation between metallicity and the Galactocentric distance.  This figure shows that clusters roughly between the Centaurus and Perseus arms have metallicities between +0.0 and +0.3 dex, while clusters along the Perseus arm display metallicities varying from -0.4 to +0.0 dex.  In particular, the two more distant clusters, Tombaugh 2 and Berkeley 20, exhibit lower metallicities ([Fe/H] $=$ -0.35 and -0.44 dex, respectively).
 
\begin{figure*}
\centering
\includegraphics[width=1.0\textwidth]{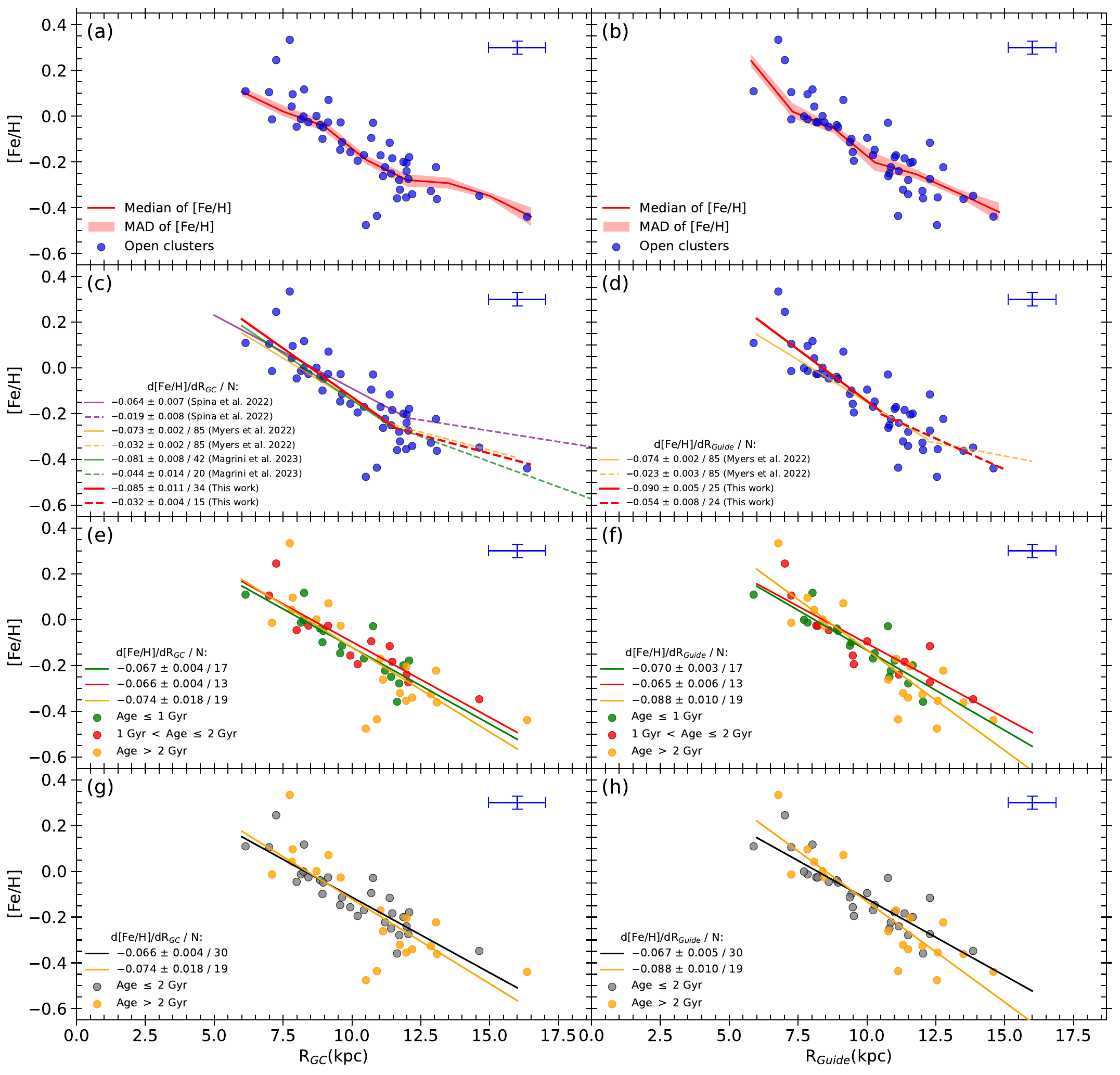}
\caption{Metallicity gradient, where R$_{GC}$ is the projected Galactocentric distance and the R$_{Guide}$ is obtained from the average between apocenter and pericenter. Blue circles represent the 49 OCs of this work. In panels (a) and (b), the red lines represent the median of [Fe/H] for each 1.5 kpc range: 6 -- 16.5 kpcs (panel a) and 6 -- 15 kpcs (panel b); the pink strip represents the median absolute deviation (MAD) of [Fe/H]. In panels (c) and (d), the red lines were obtained with two linear regressions: in panel (c), the first in the 6 -- 11.5 kpc interval (solid line) and the second in the 11.5 -- 16.5 kpc interval (dashed line); in panel (d), the first in the 6 -- 10.5 kpc interval (solid line) and the second in the 10.5 -- 15 kpc interval (dashed line). In panels (c) and (d), we compare our results with the \citet[][purple lines]{Spina2022}, \citet[][orange lines]{Myers2022}, and \citet[][green lines]{Magrini2023} results. Panels (e) and (f) separate the OCs by age: 17 OCs aged less than 1 Gyr (green circles), 13 OCs with ages between 1 and 2 Gyr (red circles), and 19 OCs aged greater than 2 Gyr (orange circles). The lines in panels (e) and (f) are the linear regressions for each of these groups of OCs: green line (OCs aged less than 1 Gyr), red line (OCs with ages between 1 and 2 Gyr), and orange line (OCs aged greater than 2 Gyr). In panels (g) and (h), all OCs of ages under 2 Gyr are represented by gray circles, while the black line is the linear regression of these OCs. The OCs ages were obtained from \cite{Cantat2020} and are aged 0.02 to 4.00 Gyr. The lines (linear regressions) this work in panels (c), (d), (e), (f), (g), and (h) were determined with the sklearn.linear\_model package in Python \citep{Pedregosa2011}. Ages were obtained from \cite{Cantat2020}.}
\label{fig:gradient}
\end{figure*}

Figure \ref{fig:gradient} shows the metallicity gradients as functions of the projected Galactocentric distance (R$_{GC}$, left panels) and the guiding center radius (R$_{Guide}$, right panels), whose values are presented in Table \ref{tab:dynamics}.  In panels (c) and (d) of Fig. \ref{fig:gradient}, two linear regressions are shown for each panel. \footnote{The lines (linear regressions) from this work in the panels (c), (d), (e), (f), (g), and (h) of Figure \ref{fig:gradient} were determined with the sklearn.linear\_model package in Python.} We compare our results with those of \citet[][purple lines in panel c]{Spina2022}, \citet[][orange lines in panels c and d]{Myers2022} and \citet[][green lines in panel c]{Magrini2023}. As found by \cite{Spina2022}, \cite{Myers2022} and \cite{Magrini2023}, our linear fits show a pronounced decrease in [Fe/H] with R$_{GC}$ and R$_{Guide}$ for R$_{GC}$ in the range 6.0--11.5 kpcs (d[Fe/H]/dR$_{GC}$ = -0.085$\pm$0.011 dex/kpc) and R$_{Guide}$ in the range 6.0--10.5 kpcs (d[Fe/H]/dR$_{Guide}$ = -0.090$\pm$0.005 dex/kpc). In panel (c), the clusters farthest from the Galactic center (R$_{GC}$ between 11.5 and 16.5 kpc) follow a shallower slope, with the same value as found by \cite{Myers2022}, with d[Fe/H]/dR$_{GC}$ = -0.032 dex/kpc, while \cite{Magrini2023} finds a slightly steeper slope (d[Fe/H]/dR$_{GC}$ = -0.044 dex/kpc), but still within our respective uncertainties. In panel (d) of Fig. \ref{fig:gradient}, we find a difference in the cutoff R$_{Guide}$ between the two groups of OCs with different slopes compared to \cite{Myers2022}. The inner and outer gradients of \cite{Myers2022} have a cutoff R$_{Guide}$ = 12.2 kpc, while in this work, the cutoff R$_{Guide}$ is 10.5 kpc.

Panels (e) and (f) of Fig. \ref{fig:gradient} show three linear regressions for three age intervals within the 49 OCs (\citealt{Cantat2020} ages; Table \ref{tab:parameters}), which were also used by \citealt{Myers2022}), with ages from 0.02 to 4.00 Gyr: 17 OCs have ages less than 1 Gyr (green circles), 13 OCs have ages between 1 and 2 Gyr (red circles), and 19 OCs have ages greater than 2 Gyr (orange circles). Using this age separation, we can see that the groups of OCs represented by the green and red circles have only small differences in their gradients with R$_{GC}$ (d[Fe/H]/dR$_{GC}$ = -0.067$\pm$0.004 and -0.066$\pm$0.004 dex/kpc, respectively), and R$_{Guide}$ (d[Fe/H]/dR$_{Guide}$ = -0.070$\pm$0.003 and -0.065$\pm$0.006 dex/kpc, respectively). Meanwhile, clusters older than 2 Gyr show significantly steeper slopes, with d[Fe/H]/dR$_{GC}$ = -0.074$\pm$0.018 dex/kpc and d[Fe/H]/dR$_{Guide}$ = -0.088$\pm$0.010 dex/kpc, as represented by the orange lines in panels (e) and (f). \citet{Carraro1998}, \citet{Friel2002}, \citet{Chen2003}, \citet{Magrini2009,Magrini2023}, and \citet{Netopil2022} also found this trend of a steeper decrease in metallicity for older OCs as they fall further from the Galactic center. In this work, it is clear that there is no break in the metallicity gradient when we consider only OCs with ages less than 2 Gyr (black lines panels in g and h) with d[Fe/H]/dR$_{GC}$ = -0.066$\pm$0.004 dex/kpc and d[Fe/H]/dR$_{Guide}$ = -0.067$\pm$0.005 dex/kpc; and ages higher than 2 Gyr (panels g and h are the same as panels e and f).

Unlike the works of \cite{Magrini2023} and \cite{Palla2024}, our sample of the youngest OCs (ages < 1 Gyr) located in the inner Galaxy does not exhibit such a pronounced difference in metallicity when compared to older OCs. \cite{Magrini2023} find that the metallicities of young OCs are significantly lower than those of older OCs in the inner Galaxy. Based on these results, \cite{Palla2024} propose that a late episode of gas accretion, triggering a metal dilution -- a scenario previously suggested by \cite{Spitoni2023} for the solar neighborhood -- may be responsible for this difference. However, our results do not indicate such a marked metallicity difference between young and old OCs in the inner Galaxy. \citet{Hu2025} recently reported results consistent with our observed metallicity gradients (Panel c of Fig. \ref{fig:gradient}), however, their findings contrast with ours regarding the dependence of metallicities on the ages of OCs, as they found a shallower metallicity gradient for older OCs.

The steeper metallicity gradients for older OCs compared to younger ones can be understood in terms of the formation and evolution of the Galactic disk. As stars form and evolve in the disk, they incorporate heavier elements, or metals, which are enriched through processes such as stellar nucleosynthesis and supernova explosions. This enrichment process and Galactic dynamics influence metallicity distribution across the disk \citep{Matteuci2021}. For the old OCs, the metallicity gradient steepens due to the differing rates of star formation and chemical enrichment in different regions of the Galactic disk. Younger clusters, on the other hand, may have formed in a more homogeneous environment, leading to a shallower gradient. Studies such as those by \cite{Magrini2023} and \cite{Netopil2022} show that metallicity gradients are steeper for older clusters, particularly as they move away from the Galactic center. The combination of longer star formation histories, galactic dynamics, and enrichment processes explains why older OCs exhibit a stronger metallicity gradient than their younger counterparts.

Our metallicity gradient results (Fig. \ref{fig:gradient}) are consistent with those obtained by \cite{Myers2022}, who found that the OCs younger than 2 Gyr align with the chemodynamical models of \cite{Chiappini2009} and \cite{Minchev2013,Minchev2014}. However, for ages greater than 2 Gyr, they observed a deviation between their sample of OCs and these models.

\section{Conclusions} \label{sec:conclusions}

This work shows that RVs and metallicities are important for classifying the star members of OCs and their membership likelihood. These findings highlight the correlation between RV proximity and high membership probabilities. The observed pattern emphasizes the effectiveness of RV as a discriminator of membership, particularly for stars within $2\sigma$ of the mean RV. The consistency of this trend across different OCs reinforces the robustness of our membership assignment criteria, demonstrating its applicability to a wide range of clusters and providing a reliable approach for refining OC membership classifications.

The dynamic analysis of these OCs reveals intriguing cases of orbital dynamics, such as NGC~6705, which exhibits a chaotic configuration influenced by the Galactic bar. The banana-shaped orbit around Lagrange points $L_4$ and $L_5$, as well as the transient behavior around the CR, suggests a possible connection to moving groups such as the Hercules stream. These findings shed light on the dynamical influence of the bar and provide valuable insights into the origin and formation processes of stellar moving groups in the Galactic disk.

Our analysis reveals a clear distinction in metallicity gradients between younger and older OCs, which can be attributed to the formation and evolutionary history of the Galactic disk. Younger OCs, with ages below 2 Gyr, exhibit shallower gradients, reflecting a more homogeneous environment during their formation. In contrast, older OCs demonstrate steeper gradients, consistent with the extended star formation history and chemical enrichment processes that have shaped the Galactic disk. These results align with previous findings and highlight the complex interplay of Galactic dynamics and chemical evolution in influencing the metallicity distribution of OCs. Our study reaffirms the importance of OCs as tracers of Galactic disk evolution and provides further evidence
 that age plays a significant role in shaping metallicity gradients.

\section*{Data availability}

The full version of Table \ref{tab:stars} is available at the CDS via anonymous ftp to \href{http://cdsarc.u-strasbg.fr}{cdsarc.u-strasbg.fr} (\href{http://130.79.128.5/}{130.79.128.5}) or via \url{http://cdsweb.u-strasbg.fr/cgi-bin/qcat?J/A+A/}.

\begin{acknowledgements}
R.G. gratefully acknowledges the grant support provided by Fundação de Amparo \`a Pesquisa do Estado do Rio de Janeiro (FAPERJ) under the P\'os-Doutorado Nota 10 (PDR-10) grant number E26-205.964/2022 and ANID Fondecyt Postdoc No. 3230001.
D.S. thank the National Council for Scientific and Technological Development – CNPq process No. 404056/2021-0.
J.G.F.-T. gratefully acknowledges the grants support provided by ANID Fondecyt Iniciaci\'on No. 11220340, and from the Joint Committee ESO-Government of Chile under the agreement 2023 ORP 062/2023.
S.D. acknowledges CNPq/MCTI for grant 306859/2022-0 and FAPERJ for grant 210.688/2024.
K.C. and V.V.S. acknowledges support from the National Science Foundation through NSF grant no. AST-2206543.
V.L.-T. acknowledges a fellowship 302195/2024-6 of the PCI Program – MCTI and fellowship 152242/2024-4 of the PDJ - MCTI and CNPq.
\\
Funding for the Sloan Digital Sky Survey IV has been provided by the Alfred P. Sloan Foundation, the U.S. Department of Energy Office of Science, and the Participating Institutions.
\\
SDSS-IV acknowledges support and resources from the Center for High Performance Computing at the University of Utah. The SDSS website is \url{www.sdss4.org}.
\\
SDSS-IV is managed by the Astrophysical Research Consortium for the Participating Institutions of the SDSS Collaboration including the Brazilian Participation Group, the Carnegie Institution for Science, Carnegie Mellon University, Center for Astrophysics | Harvard \& Smithsonian, the Chilean Participation Group, the French Participation Group, Instituto de Astrof\'isica de Canarias, The Johns Hopkins University, Kavli Institute for the Physics and Mathematics of the Universe (IPMU) / University of Tokyo, the Korean Participation Group, Lawrence Berkeley National Laboratory, Leibniz Institut f\"ur Astrophysik Potsdam (AIP),  Max-Planck-Institut f\"ur Astronomie (MPIA Heidelberg), Max-Planck-Institut f\"ur Astrophysik (MPA Garching), Max-Planck-Institut f\"ur Extraterrestrische Physik (MPE), National Astronomical Observatories of China, New Mexico State University, New York University, University of Notre Dame, Observat\'orio Nacional / MCTI, The Ohio State University, Pennsylvania State University, Shanghai Astronomical Observatory, United Kingdom Participation Group, Universidad Nacional Aut\'onoma de M\'exico, University of Arizona, University of Colorado Boulder, University of Oxford, University of Portsmouth, University of Utah, University of Virginia, University of Washington, University of Wisconsin, Vanderbilt University, and Yale University.
\\
This work has made use of data from the European Space Agency (ESA) mission Gaia (\url{https://www.cosmos.esa.int/gaia}), processed by the Gaia Data Processing and Analysis Consortium (DPAC, \url{https://www.cosmos.esa.int/web/gaia/dpac/consortium}). Funding for the DPAC has been provided by national institutions, in particular the institutions
participating in the Gaia Multilateral Agreement.
\\
Surveys: APOGEE DR17 \citep[][]{Abdurro2022} and Gaia EDR3 \citep[][]{Gaia2021}.
\\
Softwares: Astropy \citep[][]{Astropy2013}, GravPot16 (\url{https://gravpot.utinam.cnrs.fr}), HDBSCAN \citep[][]{Campello2013}, Matplotlib \citep[][]{Hunter2007}, mw-plot code (\url{https://milkyway-plot.readthedocs.io}), Numpy \citep[][]{Harris2020}, Pandas \citep{Mckinney2010}, scikit-learn \citep{Pedregosa2011} and Scipy \citep[][]{Virtanen2020}.

\end{acknowledgements}

\bibliographystyle{aa}

\twocolumn

\bibliography{reference.bib}

\newpage
\onecolumn

\begin{appendix}  \label{append}

\section{Table of the stars classified as members}

\begin{table}[h]
\caption{Stars classified as members}
\label{tab:stars} 
\centering   
\begin{tabular}{ccccc}
\hline\hline          
       &           & RA[J2000] & Dec.[J2000] & probability$^{***}$ \\ 
ID$^*$ & OC$^{**}$ & (deg)     & (deg)      & (\%)                \\ 
\hline
2MASS J05195385$+$3035095 & Berkeley 17 & 79.974375 & $+$30.585985 & 100 \\
2MASS J05202118$+$3035544 & Berkeley 17 & 80.088262 & $+$30.598465 & 100 \\
2MASS J05202905$+$3032414 & Berkeley 17 & 80.121082 & $+$30.544847 & 100 \\
2MASS J05203121$+$3035067 & Berkeley 17 & 80.130056 & $+$30.585209 & 100 \\
2MASS J05203650$+$3030351 & Berkeley 17 & 80.152085 & $+$30.509773 & 77.9 \\
2MASS J05203799$+$3034414 & Berkeley 17 & 80.158321 & $+$30.578190 & 100 \\
2MASS J05204143$+$3036042 & Berkeley 17 & 80.172651 & $+$30.601170 & 90.9 \\
2MASS J05204488$+$3038020 & Berkeley 17 & 80.187027 & $+$30.633909 & 100 \\
2MASS J05211671$+$4533170 & Berkeley 18 & 80.319637 & $+$45.554745 & 88.5 \\
2MASS J05211693$+$4524226 & Berkeley 18 & 80.320575 & $+$45.406296 & 5.84 \\
2MASS J05214903$+$4548331 & Berkeley 18 & 80.454298 & $+$45.809196 & 16.8 \\
2MASS J05214927$+$4525225 & Berkeley 18 & 80.455304 & $+$45.422939 & 100 \\
2MASS J05215476$+$4526226 & Berkeley 18 & 80.478189 & $+$45.439617 & 79.7 \\
2MASS J05215704$+$4521220 & Berkeley 18 & 80.487685 & $+$45.356113 & 100 \\
2MASS J05220382$+$4530273 & Berkeley 18 & 80.515940 & $+$45.507591 & 70.9 \\
2MASS J05220607$+$4520585 & Berkeley 18 & 80.525311 & $+$45.349609 & 100 \\
2MASS J05220733$+$4524235 & Berkeley 18 & 80.530573 & $+$45.406536 & 48.7 \\
2MASS J05220741$+$4525388 & Berkeley 18 & 80.530896 & $+$45.427448 & 47.4 \\
2MASS J05221065$+$4528494 & Berkeley 18 & 80.544413 & $+$45.480389 & 47.7 \\
2MASS J05221114$+$4517206 & Berkeley 18 & 80.546446 & $+$45.289062 & 100 \\
2MASS J05221426$+$4527000 & Berkeley 18 & 80.559428 & $+$45.450027 & 13.7 \\
2MASS J05221711$+$4523410 & Berkeley 18 & 80.571305 & $+$45.394733 & 25.4 \\
2MASS J05221874$+$4525191 & Berkeley 18 & 80.578115 & $+$45.421982 & 100 \\
2MASS J05221919$+$4529451 & Berkeley 18 & 80.579991 & $+$45.495884 & 47.7 \\
2MASS J05222163$+$4531589 & Berkeley 18 & 80.590152 & $+$45.533039 & 100 \\
2MASS J05222297$+$4518588 & Berkeley 18 & 80.595725 & $+$45.316360 & 100 \\
2MASS J05222413$+$4522021 & Berkeley 18 & 80.600544 & $+$45.367256 & 100 \\
2MASS J05222556$+$4525370 & Berkeley 18 & 80.606519 & $+$45.426964 & 45.4 \\
2MASS J05222722$+$4520061 & Berkeley 18 & 80.613418 & $+$45.335052 & 46.6 \\
2MASS J05222848$+$4523173 & Berkeley 18 & 80.618708 & $+$45.388153 & 100 \\
2MASS J05222878$+$4527249 & Berkeley 18 & 80.619922 & $+$45.456917 & 96.7 \\
2MASS J05223463$+$4531085 & Berkeley 18 & 80.644327 & $+$45.519051 & 100 \\
2MASS J05223696$+$4524397 & Berkeley 18 & 80.654019 & $+$45.411034 & 100 \\
2MASS J05223884$+$4520031 & Berkeley 18 & 80.661871 & $+$45.334213 & 47.4 \\
2MASS J05224064$+$4523367 & Berkeley 18 & 80.669348 & $+$45.393555 & 100 \\
2MASS J05224234$+$4459401 & Berkeley 18 & 80.676453 & $+$44.994488 & 36.2 \\
2MASS J05225704$+$4529067 & Berkeley 18 & 80.737684 & $+$45.485218 & 100 \\
2MASS J05230111$+$4526218 & Berkeley 18 & 80.754631 & $+$45.439396 & 100 \\
2MASS J05230556$+$4522198 & Berkeley 18 & 80.773168 & $+$45.372177 & 56.7 \\
...                       & ...         & ...       & ...          & ... \\
2MASS J12400260$-$6039545 & Trumpler 20 & 190.01086 & $-$60.665142 & 100 \\
2MASS J12400451$-$6036566 & Trumpler 20 & 190.01879 & $-$60.615726 & 100 \\
2MASS J12400755$-$6035445 & Trumpler 20 & 190.03148 & $-$60.595695 & 100 \\
2MASS J12402228$-$6037419 & Trumpler 20 & 190.09286 & $-$60.628311 & 100 \\
2MASS J12402480$-$6043101 & Trumpler 20 & 190.10335 & $-$60.719498 & 100 \\
2MASS J12402949$-$6038518 & Trumpler 20 & 190.12290 & $-$60.647732 & 100 \\
\hline
\end{tabular}
\begin{tablenotes}
\small
\item Notes: The full list of 1987 stars is available in the CDS database (Vizier). * The ID column identifies the member stars with their 2MASS names. ** The OC column identifies the OC to which the star was classified as a member. *** In the last column is the probability of star belonging to the OC using HDBSCAN.
\end{tablenotes}
\end{table}

\begin{landscape}

\section{Open cluster tables}

\begin{ThreePartTable}
 \begin{TableNotes}
    \item Notes: N$_{all}$, N$_{>50\%}$ and  N$_{>80\%}$ are the number of stars considering all probabilities, probabilities above 50 \% and probabilities above 80 \%, respectively. * Average values obtained with the stars member with probabilities greater than 80 \%. The values that are between brackets are standard deviations. The m$_{clSize}$ were chosen based on the minor standard deviation of $<$[Fe/H]$>$. ** Ages from \cite{Cantat2020}. 
\end{TableNotes}
\begin{longtable}{lcccccccccccc}
\caption{Parameters used in HDBSCAN} \\
\label{tab:parameters} \\
\hline\hline
              &           &             &             & $<$RA$>^*$ & $<$Dec.$>^*$ & $<\mu_{RA}>^*$ & $<\mu_{Dec}>^*$ & $<\pi>^*$ & $<$RV$>^*$ & $<$[Fe/H]$>^*$ &              & age$^{**}$ \\
OC            & N$_{all}$ & N$_{>50\%}$ & N$_{>80\%}$ & ($^\circ$) & ($^\circ$)  & (mas$/$yr)     & (mas$/$yr)      & (mas)     & (km/s)     & (dex)          & m$_{clSize}$ & (Gyr)      \\
\hline
\endfirsthead
\caption{Parameters used in HDBSCAN (continued)}\\
\hline\hline 
              &           &             &             & $<$RA$>^*$ & $<$Dec.$>^*$ & $<\mu_{RA}>^*$ & $<\mu_{Dec}>^*$ & $<\pi>^*$ & $<$RV$>^*$ & $<$[Fe/H]$>^*$ &              & age$^{**}$ \\
OC            & N$_{all}$ & N$_{>50\%}$ & N$_{>80\%}$ & ($^\circ$) & ($^\circ$) & (mas$/$yr)      & (mas$/$yr)      & (mas)     & (km/s)     & (dex)          & m$_{clSize}$ & (Gyr)      \\
\hline
\endhead
\hline
\insertTableNotes
\endfoot
Berkeley 17   & 8   & 8   & 7   & $+$80.1[0.07] & $+$30.6[0.03] & $+$2.53[0.04] & $-$0.39[0.07] & 0.30[0.02] & $-$73.4[0.27] & $-$0.17[0.03] & 6  & 7.24 \\ 
Berkeley 18   & 31  & 19  & 16  & $+$80.6[0.10] & $+$45.4[0.07] & $+$0.76[0.05] & $-$0.09[0.05] & 0.16[0.03] & $-$2.58[0.33] & $-$0.36[0.03] & 3  & 4.37 \\ 
Berkeley 19   & 6   & 5   & 3   & $+$81.0[0.03] & $+$29.8[0.29] & $+$0.48[0.23] & $-$0.10[0.18] & 0.26[0.13] & $+$18.1[0.57] & $-$0.36[0.02] & 2  & 2.19 \\
Berkeley 20   & 11  & 9   & 4   & $+$83.2[0.05] & $+$0.19[0.01] & $+$0.79[0.15] & $-$0.30[0.07] & 0.06[0.04] & $+$78.6[2.00] & $-$0.44[0.06] & 3  & 4.79 \\
Berkeley 21   & 10  & 8   & 8   & $+$88.0[0.06] & $+$21.8[0.14] & $+$0.50[0.09] & $-$0.82[0.53] & 0.16[0.05] & $+$0.84[0.55] & $-$0.22[0.04] & 6  & 2.14 \\
Berkeley 22   & 9   & 5   & 3   & $+$89.6[0.05] & $+$7.76[0.03] & $+$0.61[0.01] & $-$0.37[0.02] & 0.15[0.04] & $+$95.3[0.13] & $-$0.33[0.02] & 3  & 2.45 \\
Berkeley 33   & 18  & 9   & 4   & $+$104.[0.05] & $-$13.0[0.29] & $-$0.79[0.17] & $+$1.47[0.23] & 0.19[0.03] & $+$78.2[0.34] & $-$0.28[0.04] & 3  & 0.23 \\
Berkeley 53   & 14  & 8   & 5   & $+$314.[0.04] & $+$51.1[0.04] & $-$3.83[0.09] & $-$5.69[0.04] & 0.23[0.03] & $-$36.2[0.47] & $-$0.10[0.03] & 4  & 0.98 \\
Berkeley 66   & 11  & 7   & 6   & $+$46.1[0.04] & $+$58.8[0.02] & $-$0.17[0.07] & $+$0.07[0.05] & 0.18[0.06] & $-$50.0[0.23] & $-$0.20[0.02] & 2  & 3.09 \\
Berkeley 71   & 10  & 4   & 4   & $+$85.2[0.02] & $+$32.3[0.02] & $+$0.65[0.03] & $-$1.67[0.01] & 0.26[0.04] & $-$9.44[0.12] & $-$0.25[0.02] & 3  & 0.87 \\
Berkeley 98   & 4   & 3   & 3   & $+$341.[0.44] & $+$52.6[0.14] & $-$1.32[0.01] & $-$3.23[0.06] & 0.25[0.01] & $-$69.1[0.26] & $-$0.03[0.02] & 2  & 2.45 \\
Collinder 261 & 8   & 5   & 4   & $+$190.[0.05] & $-$68.4[0.02] & $-$6.42[0.08] & $-$2.63[0.04] & 0.35[0.01] & $-$25.2[0.08] & $-$0.01[0.01] & 3  & 6.31 \\
Czernik 20    & 20  & 12  & 7   & $+$80.0[0.25] & $+$39.4[0.27] & $+$0.44[0.22] & $-$1.60[0.25] & 0.27[0.06] & $+$30.8[0.77] & $-$0.18[0.03] & 5  & 1.66 \\
Czernik 21    & 4   & 4   & 3   & $+$81.7[0.07] & $+$36.1[0.06] & $+$2.18[0.13] & $-$1.01[0.08] & 0.24[0.02] & $+$45.5[0.57] & $-$0.32[0.00] & 2  & 2.57 \\
FSR2007 0494  & 6   & 3   & 3   & $+$6.43[0.05] & $+$63.8[0.02] & $-$2.49[0.04] & $-$0.93[0.03] & 0.21[0.02] & $-$64.7[0.16] & $-$0.03[0.01] & 2  & 0.89 \\
IC 166        & 25  & 9   & 7   & $+$28.1[0.09] & $+$61.8[0.05] & $-$1.46[0.05] & $+$1.05[0.04] & 0.22[0.07] & $-$39.8[0.19] & $-$0.12[0.03] & 5  & 1.32 \\
IC 1369       & 7   & 3   & 3   & $+$318.[0.04] & $+$47.8[0.01] & $-$4.56[0.02] & $-$5.70[0.07] & 0.27[0.02] & $-$48.7[0.09] & $-$0.04[0.03] & 2  & 0.29 \\
King 5        & 16  & 15  & 3   & $+$48.7[0.14] & $+$52.7[0.04] & $-$0.29[0.16] & $-$1.32[0.04] & 0.43[0.02] & $-$43.8[0.70] & $-$0.16[0.02] & 2  & 1.02 \\
King 7        & 7   & 5   & 4   & $+$59.8[0.04] & $+$51.8[0.02] & $+$0.99[0.07] & $-$1.17[0.04] & 0.37[0.04] & $-$10.4[0.16] & $-$0.17[0.02] & 2  & 0.22 \\
King 8        & 5   & 3   & 3   & $+$87.2[0.25] & $+$33.8[0.18] & $+$0.35[0.08] & $-$2.15[0.43] & 0.18[0.06] & $-$1.12[0.17] & $-$0.18[0.05] & 2  & 0.83 \\
M 35          & 101 & 76  & 32  & $+$92.3[0.15] & $+$24.4[0.17] & $+$2.23[0.13] & $-$2.80[0.13] & 1.15[0.03] & $-$7.42[0.42] & $-$0.05[0.04] & 6  & 0.15 \\
M 44          & 185 & 143 & 69  & $+$129.[0.46] & $+$19.6[0.41] & $-$35.8[0.56] & $-$12.9[0.48] & 5.42[0.07] & $+$35.3[0.51] & $+$0.12[0.07] & 10 & 0.68 \\
M 67          & 378 & 328 & 257 & $+$133.[0.17] & $+$11.8[0.19] & $-$11.0[0.15] & $-$2.92[0.16] & 1.15[0.05] & $+$34.2[0.57] & $+$0.00[0.06] & 5  & 4.27 \\
Melotte 71    & 7   & 6   & 4   & $+$114.[0.03] & $-$12.1[0.03] & $-$2.41[0.09] & $+$4.26[0.20] & 0.49[0.07] & $+$51.0[0.39] & $-$0.15[0.02] & 2  & 0.98 \\
NGC 188       & 69  & 45  & 39  & $+$12.0[1.28] & $+$85.2[0.16] & $-$2.34[0.10] & $-$1.04[0.07] & 0.52[0.02] & $-$41.8[0.63] & $+$0.07[0.04] & 3  & 7.08 \\
NGC 752       & 105 & 76  & 35  & $+$29.3[0.28] & $+$37.8[0.19] & $+$9.76[0.18] & $-$11.8[0.19] & 2.26[0.05] & $+$6.21[0.41] & $-$0.03[0.05] & 6  & 1.17 \\
NGC 1193      & 9   & 4   & 4   & $+$46.6[0.03] & $+$44.4[0.02] & $-$0.24[0.05] & $-$0.43[0.04] & 0.18[0.03] & $-$84.8[0.24] & $-$0.34[0.02] & 3  & 5.13 \\
NGC 1245      & 28  & 22  & 20  & $+$48.7[0.10] & $+$47.3[0.06] & $+$0.47[0.07] & $-$1.67[0.04] & 0.30[0.02] & $-$29.4[0.43] & $-$0.10[0.02] & 5  & 1.20 \\
NGC 1798      & 9   & 7   & 7   & $+$77.9[0.03] & $+$47.7[0.02] & $+$0.76[0.07] & $-$0.37[0.04] & 0.21[0.03] & $+$2.73[0.58] & $-$0.27[0.03] & 2  & 1.66 \\
NGC 1857      & 5   & 5   & 3   & $+$80.0[0.28] & $+$39.7[0.24] & $+$0.38[0.17] & $-$1.55[0.27] & 0.31[0.10] & $+$1.38[0.17] & $-$0.22[0.06] & 2  & 0.25 \\
NGC 1907      & 4   & 3   & 3   & $+$82.0[0.04] & $+$35.3[0.01] & $-$0.18[0.04] & $-$3.45[0.07] & 0.64[0.04] & $+$2.68[0.16] & $-$0.11[0.01] & 2  & 0.59 \\
NGC 2158      & 58  & 57  & 56  & $+$91.9[0.06] & $+$24.1[0.08] & $-$0.23[0.09] & $-$1.99[0.07] & 0.23[0.05] & $+$27.2[1.24] & $-$0.24[0.03] & 3  & 1.55 \\
NGC 2204      & 24  & 16  & 9   & $+$93.8[0.10] & $-$18.6[0.06] & $-$0.58[0.04] & $+$1.95[0.04] & 0.22[0.02] & $+$92.2[0.11] & $-$0.26[0.04] & 5  & 2.09 \\
NGC 2243      & 15  & 13  & 13  & $+$97.4[0.07] & $-$31.3[0.10] & $-$1.24[0.04] & $+$5.50[0.04] & 0.22[0.01] & $+$60.0[0.45] & $-$0.48[0.04] & 2  & 4.37 \\
NGC 2324      & 8   & 4   & 4   & $+$106.[0.03] & $+$1.05[0.06] & $-$0.36[0.02] & $-$0.05[0.05] & 0.22[0.01] & $+$42.4[0.18] & $-$0.20[0.02] & 3  & 0.54 \\
NGC 2420      & 19  & 19  & 16  & $+$115.[0.03] & $+$21.6[0.05] & $-$1.25[0.04] & $-$1.99[0.07] & 0.41[0.02] & $+$74.4[0.30] & $-$0.20[0.02] & 2  & 1.74 \\
NGC 4337      & 12  & 7   & 4   & $+$186.[0.03] & $-$58.1[0.02] & $-$8.82[0.08] & $+$1.55[0.03] & 0.38[0.01] & $-$17.7[0.16] & $+$0.25[0.04] & 3  & 1.45 \\
NGC 6705      & 20  & 7   & 6   & $+$283.[0.02] & $-$6.27[0.04] & $-$1.49[0.11] & $-$4.20[0.11] & 0.41[0.03] & $+$34.8[0.20] & $+$0.11[0.04] & 3  & 0.31 \\
NGC 6791      & 79  & 64  & 47  & $+$290.[0.08] & $+$37.8[0.08] & $-$0.39[0.06] & $-$2.25[0.06] & 0.21[0.02] & $-$47.2[0.77] & $+$0.33[0.04] & 6  & 6.31 \\
NGC 6811      & 11  & 6   & 3   & $+$294.[0.07] & $+$46.5[0.06] & $-$3.29[0.06] & $-$8.83[0.06] & 0.89[0.00] & $+$7.40[0.06] & $-$0.05[0.01] & 2  & 1.07 \\
NGC 6819      & 79  & 39  & 19  & $+$295.[0.05] & $+$40.2[0.06] & $-$2.85[0.10] & $-$3.94[0.07] & 0.38[0.02] & $+$2.76[0.38] & $+$0.04[0.03] & 5  & 2.24 \\
NGC 7058      & 7   & 6   & 5   & $+$321.[0.19] & $+$50.9[0.07] & $+$7.40[0.18] & $+$2.96[0.37] & 2.73[0.02] & $-$19.2[0.21] & $-$0.01[0.03] & 4  & 0.04 \\
NGC 7789      & 95  & 62  & 49  & $+$359.[0.16] & $+$56.8[0.10] & $-$0.94[0.11] & $-$2.02[0.10] & 0.50[0.02] & $-$54.4[0.71] & $-$0.03[0.03] & 5  & 1.55 \\
Pleiades      & 306 & 223 & 86  & $+$56.5[0.59] & $+$24.1[0.63] & $+$20.0[0.69] & $-$45.5[0.71] & 7.38[0.10] & $+$5.95[0.60] & $-$0.00[0.06] & 5  & 0.08 \\
Ruprecht 147  & 59  & 30  & 16  & $+$289.[0.26] & $-$16.3[0.20] & $-$0.80[0.31] & $-$26.8[0.28] & 3.28[0.03] & $+$42.2[0.28] & $+$0.10[0.02] & 3  & 3.02 \\
Teutsch 51    & 10  & 5   & 2   & $+$88.5[0.03] & $+$26.8[0.01] & $+$0.56[0.05] & $-$0.22[0.10] & 0.24[0.03] & $+$17.7[0.20] & $-$0.36[0.01] & 2  & 0.68 \\
Tombaugh 2    & 17  & 9   & 5   & $+$106.[0.02] & $-$20.8[0.02] & $-$0.51[0.04] & $+$1.47[0.08] & 0.07[0.07] & $+$121.[0.34] & $-$0.35[0.02] & 4  & 1.62 \\
Trumpler 5    & 12  & 5   & 5   & $+$99.3[0.11] & $+$9.54[0.15] & $-$0.65[0.08] & $+$0.24[0.07] & 0.30[0.01] & $+$50.5[0.42] & $-$0.44[0.02] & 3  & 4.27 \\
Trumpler 20   & 26  & 25  & 23  & $+$190.[0.11] & $-$60.6[0.05] & $-$7.08[0.07] & $+$0.16[0.11] & 0.27[0.02] & $-$40.1[0.47] & $+$0.11[0.02] & 2  & 1.86 \\
\end{longtable}
\end{ThreePartTable}

\begin{longtable}{lccccccccccc}
\caption{Galactocentric positions and dynamic properties of OCs} \\
\label{tab:dynamics} \\
\hline\hline
                 & X     & Y        & Z        & R$_{GC}$ & V$_{R}$             & V$\phi$ &                 & Z$_{max}$       & pericenter      & apocenter       & R$_{Guide}$     \\
OC               & (kpc) & (kpc)    & (kpc)    & (kpc)    & (km$\cdot$s$^{-1}$) & (km$\cdot$s$^{-1}$) & e   & (kpc)           & (kpc)           & (kpc)           & (kpc)           \\ 
\hline
\endfirsthead
\caption{Galactocentric positions and dynamic properties of OCs (continued)}\\
\hline\hline
                 & X     & Y        & Z        & R$_{GC}$ & V$_{R}$             & V$\phi$ &                 & Z$_{max}$       & pericenter      & apocenter       & R$_{Guide}$     \\
OC               & (kpc) & (kpc)    & (kpc)    & (kpc)    & (km$\cdot$s$^{-1}$) & (km$\cdot$s$^{-1}$) & e   & (kpc)           & (kpc)           & (kpc)           & (kpc)           \\ 
\hline
\endhead
\hline
\endfoot
Berkeley 17      & 10.91 & $-$0.222 & $-$0.188 & 11.04 & $-$76.02              & 228.5   & 0.239$\pm$0.001 & 1.578$\pm$0.105 & 8.412$\pm$0.106 & 13.70$\pm$0.168 & 11.05$\pm$0.273 \\
Berkeley 18      & 12.89 & $-$1.435 & $+$0.450 & 13.09 & $+$15.75              & 243.8   & 0.065$\pm$0.007 & 1.039$\pm$0.259 & 12.63$\pm$0.683 & 14.38$\pm$0.971 & 13.50$\pm$1.654 \\
Berkeley 19      & 11.84 & $-$0.220 & $-$0.235 & 11.97 & $+$12.33              & 251.1   & 0.073$\pm$0.013 & 0.506$\pm$0.233 & 11.68$\pm$1.188 & 13.46$\pm$1.517 & 12.57$\pm$2.705 \\
Berkeley 20      & 15.87 & $+$3.419 & $-$2.684 & 16.36 & $+$10.17              & 205.9   & 0.130$\pm$0.025 & 2.769$\pm$0.846 & 12.70$\pm$1.353 & 16.49$\pm$2.274 & 14.59$\pm$3.627 \\
Berkeley 21      & 12.92 & $+$0.593 & $-$0.216 & 13.05 & $-$23.69              & 233.4   & 0.083$\pm$0.017 & 0.371$\pm$0.201 & 11.64$\pm$0.919 & 13.90$\pm$1.045 & 12.77$\pm$1.964 \\
Berkeley 22      & 12.63 & $+$1.674 & $-$0.700 & 12.87 & $+$45.47              & 218.3   & 0.166$\pm$0.005 & 0.831$\pm$0.094 & 10.02$\pm$0.516 & 14.01$\pm$0.569 & 12.01$\pm$1.085 \\
Berkeley 33      & 11.15 & $+$3.182 & $-$0.358 & 11.72 & $+$4.108              & 236.6   & 0.028$\pm$0.014 & 0.379$\pm$0.042 & 11.23$\pm$0.625 & 11.76$\pm$0.412 & 11.50$\pm$1.037 \\
Berkeley 53      & 8.020 & $-$3.668 & $+$0.240 & 8.929 & $-$25.84              & 256.1   & 0.100$\pm$0.036 & 0.329$\pm$0.080 & 8.496$\pm$0.222 & 10.38$\pm$1.118 & 9.437$\pm$1.340 \\
Berkeley 66      & 11.48 & $-$2.983 & $+$0.016 & 11.98 & $+$7.051              & 233.0   & 0.038$\pm$0.014 & 0.198$\pm$0.045 & 11.20$\pm$0.886 & 11.96$\pm$1.100 & 11.58$\pm$1.986 \\
Berkeley 71      & 11.30 & $-$0.194 & $+$0.050 & 11.42 & $-$14.89              & 228.9   & 0.064$\pm$0.008 & 0.075$\pm$0.010 & 10.12$\pm$0.171 & 11.51$\pm$0.378 & 10.82$\pm$0.550 \\
Berkeley 98      & 8.850 & $-$3.380 & $-$0.342 & 9.588 & $+$3.104              & 210.2   & 0.145$\pm$0.005 & 0.688$\pm$0.068 & 7.234$\pm$0.127 & 9.681$\pm$0.080 & 8.458$\pm$0.207 \\
Collinder 261    & 6.605 & $+$2.261 & $-$0.257 & 7.097 & $-$9.867              & 249.7   & 0.061$\pm$0.002 & 0.537$\pm$0.020 & 6.812$\pm$0.032 & 7.689$\pm$0.034 & 7.251$\pm$0.066 \\
Czernik 20       & 11.32 & $-$0.685 & $+$0.073 & 11.46 & $+$38.67              & 236.6   & 0.124$\pm$0.009 & 0.118$\pm$0.063 & 9.937$\pm$0.335 & 12.80$\pm$0.567 & 11.37$\pm$0.902 \\
Czernik 21       & 11.61 & $-$0.516 & $+$0.032 & 11.74 & $+$48.81              & 226.1   & 0.161$\pm$0.006 & 1.235$\pm$0.149 & 9.481$\pm$0.126 & 13.14$\pm$0.256 & 11.31$\pm$0.382 \\
FSR2007 0494     & 10.05 & $-$3.535 & $+$0.073 & 10.77 & $-$6.635              & 241.9   & 0.031$\pm$0.004 & 0.195$\pm$0.030 & 10.45$\pm$0.311 & 11.07$\pm$0.290 & 10.76$\pm$0.602 \\
IC 166           & 10.76 & $-$3.287 & $-$0.014 & 11.37 & $+$12.59              & 255.5   & 0.093$\pm$0.047 & 0.755$\pm$0.385 & 11.18$\pm$1.066 & 13.39$\pm$2.651 & 12.29$\pm$3.717 \\
IC 1369          & 7.974 & $-$3.570 & $-$0.025 & 8.848 & $-$37.53              & 243.7   & 0.118$\pm$0.007 & 0.054$\pm$0.027 & 7.861$\pm$0.327 & 9.944$\pm$0.554 & 8.902$\pm$0.881 \\
King 5           & 9.736 & $-$1.272 & $-$0.159 & 9.940 & $-$14.62              & 230.5   & 0.091$\pm$0.002 & 0.179$\pm$0.011 & 8.623$\pm$0.071 & 10.34$\pm$0.089 & 9.483$\pm$0.159 \\
King 7           & 10.23 & $-$1.299 & $-$0.045 & 10.43 & $+$18.93              & 234.9   & 0.084$\pm$0.018 & 0.120$\pm$0.014 & 9.379$\pm$0.068 & 11.06$\pm$0.376 & 10.22$\pm$0.443 \\
King 8           & 11.95 & $-$0.262 & $+$0.210 & 12.08 & $-$4.102              & 218.8   & 0.085$\pm$0.035 & 0.323$\pm$0.140 & 10.03$\pm$0.619 & 11.98$\pm$0.470 & 11.00$\pm$1.089 \\
M 35             & 8.834 & $+$0.097 & $+$0.034 & 8.956 & $-$22.84              & 243.4   & 0.080$\pm$0.001 & 0.190$\pm$0.011 & 8.234$\pm$0.020 & 9.675$\pm$0.022 & 8.955$\pm$0.042 \\
M 44             & 8.140 & $+$0.067 & $+$0.096 & 8.262 & $+$29.82              & 236.8   & 0.099$\pm$0.004 & 0.110$\pm$0.004 & 7.229$\pm$0.020 & 8.817$\pm$0.097 & 8.023$\pm$0.118 \\
M 67             & 8.581 & $+$0.417 & $+$0.445 & 8.772 & $+$24.89              & 233.8   & 0.086$\pm$0.002 & 0.557$\pm$0.041 & 7.663$\pm$0.032 & 9.110$\pm$0.046 & 8.387$\pm$0.078 \\
Melotte 71       & 9.329 & $+$1.527 & $+$0.160 & 9.573 & $+$16.81              & 255.4   & 0.107$\pm$0.027 & 0.323$\pm$0.040 & 9.104$\pm$0.279 & 11.45$\pm$0.682 & 10.28$\pm$0.961 \\
NGC 188          & 8.917 & $-$1.420 & $+$0.695 & 9.148 & $-$9.340              & 242.6   & 0.045$\pm$0.003 & 0.847$\pm$0.028 & 8.730$\pm$0.070 & 9.556$\pm$0.066 & 9.143$\pm$0.136 \\
NGC 752          & 8.293 & $-$0.273 & $-$0.172 & 8.420 & $+$13.47              & 236.4   & 0.060$\pm$0.002 & 0.273$\pm$0.005 & 7.675$\pm$0.021 & 8.648$\pm$0.009 & 8.161$\pm$0.030 \\
NGC 1193         & 11.81 & $-$2.487 & $-$0.978 & 12.19 & $-$37.28              & 222.5   & 0.136$\pm$0.008 & 1.183$\pm$0.038 & 9.936$\pm$0.442 & 13.05$\pm$0.373 & 11.49$\pm$0.815 \\
NGC 1245         & 10.46 & $-$1.617 & $-$0.461 & 10.70 & $+$7.018              & 225.8   & 0.067$\pm$0.011 & 0.477$\pm$0.019 & 9.333$\pm$0.150 & 10.68$\pm$0.110 & 10.01$\pm$0.260 \\
NGC 1798         & 11.85 & $-$1.348 & $+$0.346 & 12.05 & $+$23.22              & 242.6   & 0.080$\pm$0.005 & 0.676$\pm$0.060 & 11.31$\pm$0.220 & 13.27$\pm$0.375 & 12.29$\pm$0.594 \\
NGC 1857         & 11.06 & $-$0.643 & $+$0.078 & 11.20 & $+$8.661              & 235.4   & 0.044$\pm$0.019 & 0.106$\pm$0.056 & 10.40$\pm$0.571 & 11.33$\pm$0.704 & 10.87$\pm$1.275 \\
NGC 1907         & 9.514 & $-$0.196 & $+$0.008 & 9.638 & $-$0.890              & 237.3   & 0.043$\pm$0.004 & 0.161$\pm$0.019 & 8.972$\pm$0.027 & 9.784$\pm$0.083 & 9.378$\pm$0.111 \\
NGC 2158         & 11.85 & $+$0.448 & $+$0.121 & 11.98 & $+$4.670              & 224.1   & 0.065$\pm$0.018 & 0.446$\pm$0.239 & 10.42$\pm$0.317 & 11.89$\pm$0.773 & 11.16$\pm$1.090 \\
NGC 2204         & 10.66 & $+$2.751 & $-$1.107 & 11.13 & $+$21.66              & 231.4   & 0.087$\pm$0.010 & 1.228$\pm$0.062 & 9.835$\pm$0.281 & 11.72$\pm$0.116 & 10.78$\pm$0.397 \\
NGC 2243         & 9.882 & $+$3.192 & $-$1.207 & 10.50 & $+$25.12              & 276.9   & 0.184$\pm$0.016 & 1.805$\pm$0.146 & 10.23$\pm$0.149 & 14.85$\pm$0.717 & 12.54$\pm$0.865 \\
NGC 2324         & 11.52 & $+$2.322 & $+$0.241 & 11.87 & $-$21.14              & 235.7   & 0.066$\pm$0.009 & 0.277$\pm$0.023 & 10.89$\pm$0.275 & 12.43$\pm$0.526 & 11.66$\pm$0.801 \\
NGC 2420         & 10.06 & $+$0.674 & $+$0.774 & 10.20 & $+$42.75              & 221.9   & 0.158$\pm$0.002 & 0.905$\pm$0.029 & 8.019$\pm$0.053 & 11.03$\pm$0.076 & 9.524$\pm$0.129 \\
NGC 4337         & 6.810 & $+$2.120 & $+$0.194 & 7.249 & $+$17.87              & 238.1   & 0.069$\pm$0.004 & 0.273$\pm$0.008 & 6.536$\pm$0.056 & 7.501$\pm$0.022 & 7.019$\pm$0.078 \\
NGC 6705         & 5.919 & $-$1.075 & $-$0.115 & 6.136 & $-$22.31              & 236.7   & 0.137$\pm$0.045 & 0.124$\pm$0.007 & 4.679$\pm$0.514 & 7.090$\pm$0.507 & 5.885$\pm$1.021 \\
NGC 6791         & 6.593 & $-$3.858 & $+$0.792 & 7.742 & $+$69.41              & 191.8   & 0.294$\pm$0.010 & 1.060$\pm$0.085 & 4.784$\pm$0.055 & 8.777$\pm$0.104 & 6.781$\pm$0.158 \\
NGC 6811         & 7.801 & $-$1.051 & $+$0.229 & 7.992 & $-$25.67              & 261.7   & 0.113$\pm$0.001 & 0.279$\pm$0.002 & 7.635$\pm$0.005 & 9.587$\pm$0.010 & 8.611$\pm$0.015 \\
NGC 6819         & 7.304 & $-$2.426 & $+$0.378 & 7.812 & $+$12.21              & 251.2   & 0.063$\pm$0.003 & 0.500$\pm$0.022 & 7.576$\pm$0.017 & 8.600$\pm$0.045 & 8.088$\pm$0.062 \\
NGC 7058         & 8.019 & $-$0.361 & $+$0.004 & 8.149 & $+$11.16              & 236.7   & 0.050$\pm$0.001 & 0.027$\pm$0.007 & 7.452$\pm$0.015 & 8.241$\pm$0.008 & 7.846$\pm$0.023 \\
NGC 7789         & 8.839 & $-$1.758 & $-$0.182 & 9.133 & $-$4.104              & 216.8   & 0.122$\pm$0.003 & 0.202$\pm$0.017 & 7.208$\pm$0.077 & 9.211$\pm$0.052 & 8.209$\pm$0.129 \\
Pleiades         & 8.120 & $-$0.029 & $-$0.054 & 8.242 & $-$3.245              & 228.6   & 0.076$\pm$0.003 & 0.126$\pm$0.008 & 7.132$\pm$0.037 & 8.297$\pm$0.005 & 7.714$\pm$0.043 \\
Ruprecht 147     & 7.725 & $-$0.105 & $-$0.066 & 7.848 & $-$55.90              & 238.1   & 0.173$\pm$0.001 & 0.318$\pm$0.010 & 6.485$\pm$0.018 & 9.197$\pm$0.024 & 7.841$\pm$0.042 \\
Teutsch 51       & 11.52 & $+$0.169 & $+$0.029 & 11.64 & $+$2.474              & 248.1   & 0.046$\pm$0.007 & 0.436$\pm$0.064 & 11.49$\pm$0.370 & 12.58$\pm$0.404 & 12.04$\pm$0.774 \\
Tombaugh 2       & 12.96 & $+$6.539 & $-$0.986 & 14.62 & $+$10.10              & 225.2   & 0.087$\pm$0.073 & 0.993$\pm$0.426 & 13.08$\pm$4.105 & 14.63$\pm$3.371 & 13.86$\pm$7.475 \\
Trumpler 5       & 10.72 & $+$1.145 & $+$0.060 & 10.90 & $+$12.28              & 246.4   & 0.053$\pm$0.003 & 0.080$\pm$0.016 & 10.55$\pm$0.117 & 11.72$\pm$0.139 & 11.13$\pm$0.255 \\
Trumpler 20      & 6.281 & $+$2.808 & $+$0.128 & 6.991 & $-$0.052              & 255.3   & 0.068$\pm$0.002 & 0.144$\pm$0.010 & 6.755$\pm$0.046 & 7.750$\pm$0.076 & 7.252$\pm$0.122 \\
\end{longtable}
\end{landscape}

\section{Orbits (adopting a bar pattern speed of 31 km s$^{-1}$ kpc$^{-1}$)}

\begin{figure}[h]
\centering
\includegraphics[width=1.0\textwidth]{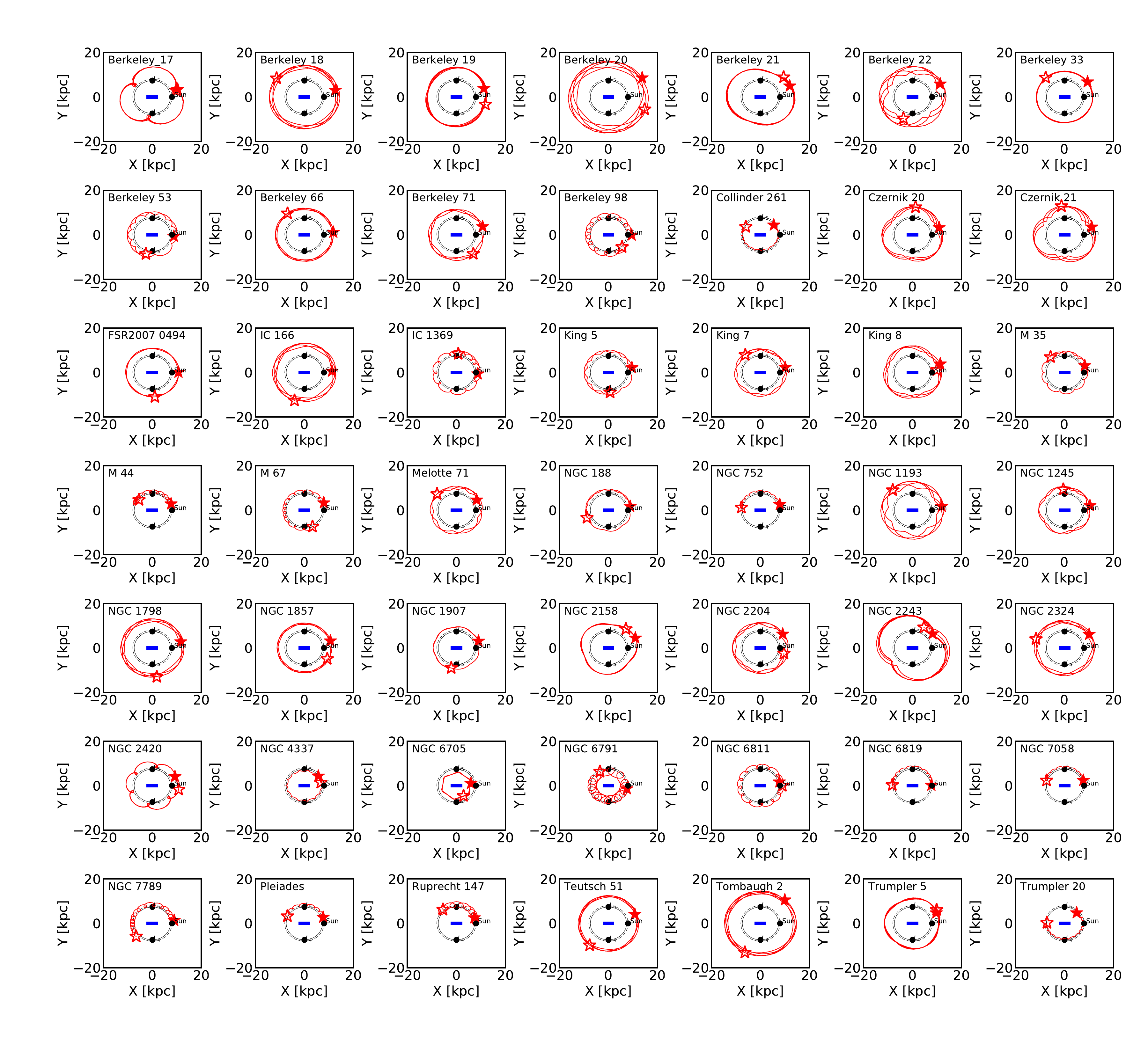}
\caption{Representative orbits for the sample of selected OCs projected on the equatorial GP in the non-inertial frame with a bar pattern speed of 31 km s$^{-1}$ kpc$^{-1}$, and time-integrated backward over 2 Gyr. The solid and dashed circles show the locations of the CR and solar orbit, respectively, while the black dots mark the positions of the Lagrange points of the Galactic bar, $L_4$ and $L_5$, and the current Sun position. The horizontal solid blue line shows the extension of the bar in our model. The solid red line shows the orbits of the selected OCs from the observables without error bars. The red filled and unfilled star symbols indicate the initial and final positions of the OCs in our simulations, respectively.
}
\label{equatorial1}
\end{figure}

\newpage

\section{Orbits (adopting a bar pattern speed of 41 km s$^{-1}$ kpc$^{-1}$)}

\begin{figure}[h]
\centering
\includegraphics[width=1.0\textwidth]{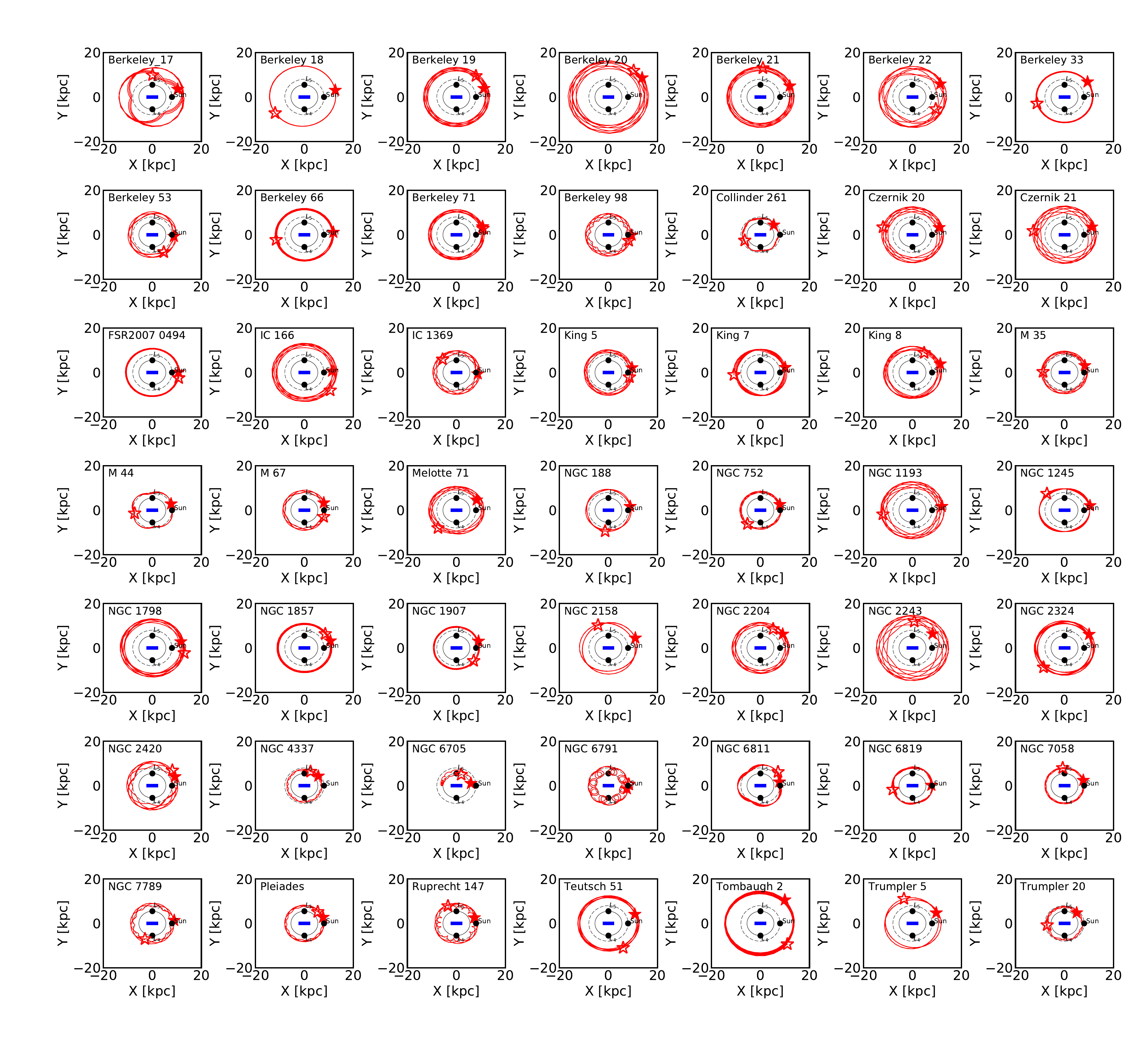}
\caption{Same as Fig. \ref{equatorial1} but adopting a bar pattern speed of 41 km s$^{-1}$ kpc$^{-1}$.}
\label{equatorial2}
\end{figure}

\newpage

\section{Orbits (adopting a bar pattern speed of 51 km s$^{-1}$ kpc$^{-1}$)}

\begin{figure}[h]
\centering
\includegraphics[width=1.0\textwidth]{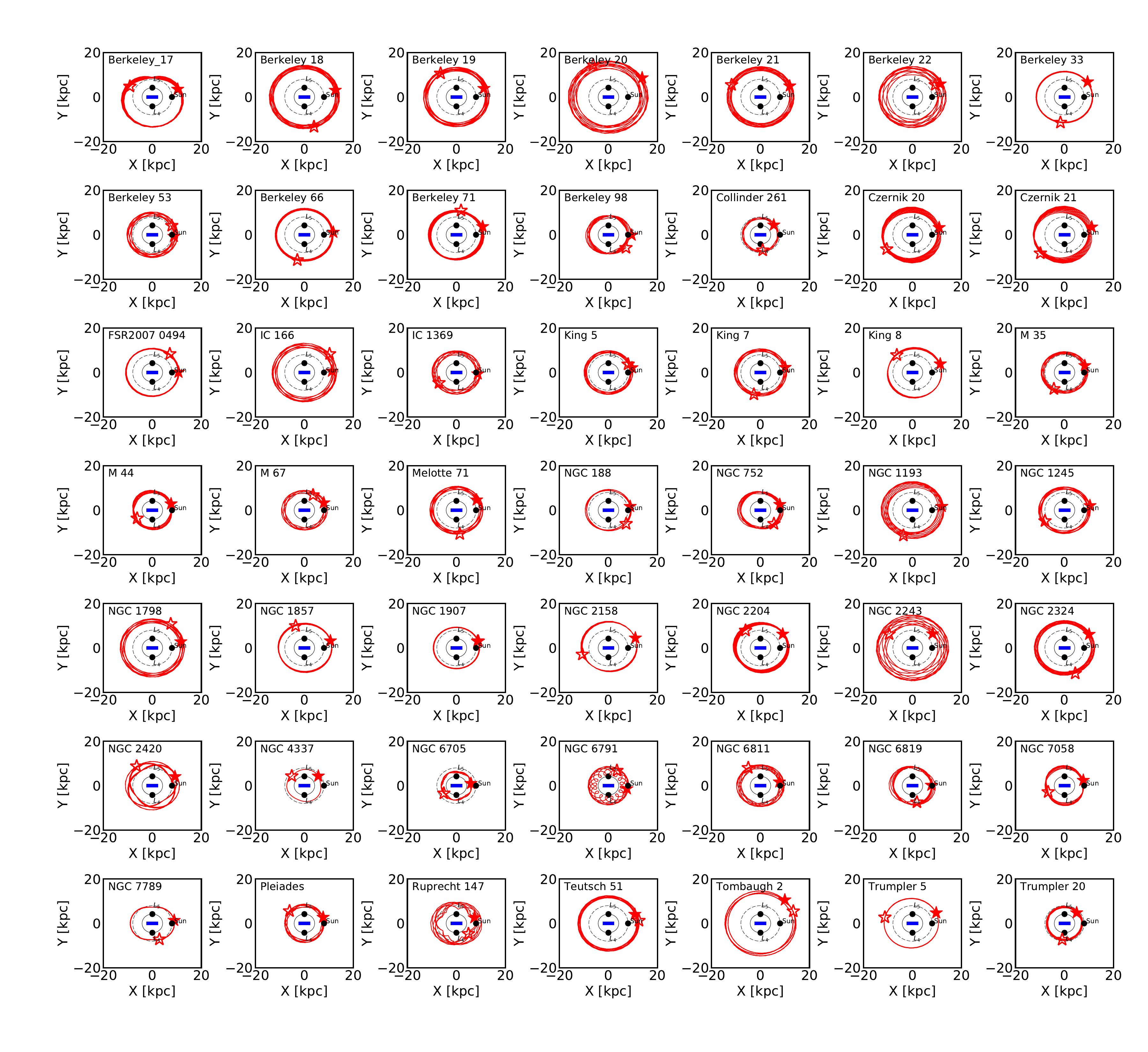}
\caption{Same as Fig. \ref{equatorial1} but adopting a bar pattern speed of 51 km s$^{-1}$ kpc$^{-1}$.}
\label{equatorial3}
\end{figure}

\end{appendix}

\end{document}